\documentclass[preprint]{aastex}
\usepackage{graphicx}
\usepackage{natbib}

\shorttitle{Astrometric Methods for Planet Searches}
\shortauthors{A. Sozzetti}

\begin{document}


\title{Astrometric Methods and Instrumentation to
Identify and Characterize Extrasolar Planets: A Review}


\author{Alessandro Sozzetti\altaffilmark{1,2}}
\altaffiltext{1}{University of Pittsburgh, Dept. of Physics \&
Astronomy, Pittsburgh, PA 15260, USA} \altaffiltext{2}{Smithsonian
Astrophysical Observatory, Harvard-Smithsonian Center for
Astrophysics, 60 Garden Street, Cambridge, MA 02138}
\email{asozzetti@cfa.harvard.edu}




\begin{abstract}
I present a review of astrometric techniques and instrumentation utilized to
search for, detect, and characterize extra-solar planets. First, I briefly summarize
the properties of the present-day sample of extrasolar planets, in connection
with predictions from theoretical models of planet formation and evolution.
Next, the generic approach to planet detection with astrometry is described,
with significant discussion of a variety of technical, statistical, and
astrophysical issues to be faced by future ground-based as well as space-borne
efforts in order to achieve the required degree of measurement precision. After
a brief summary of past and present efforts to detect planets via milli-arcsecond
astrometry, I then discuss the planet-finding capabilities of future astrometric
observatories aiming at micro-arcsecond precision. Lastly,
I outline a number experiments that can be conducted by means of high-precision
astrometry during the next decade, to illustrate its potential for important
contributions to planetary science, in comparison with other indirect and
direct methods for the detection and characterization of planetary systems.

\end{abstract}



\keywords{astrometry --- planetary systems --- stars: statistics ---
instrumentation: miscellaneous ---  methods: statistical ---  methods: numerical}


\section{Introduction}

The astrophysics of planetary systems is a good example of a branch of
science in which theory is mostly driven by observations. Hardly any of
the properties of the sample of $\sim 150$ extrasolar giant planets discovered
to-date\footnote{See for example the continuously updated
Extrasolar Planet Encyclopedia website
(\url{http://www.obspm.fr/encycl/encycl.html}), and references to
published papers therein. Objects with masses $M_p$ below the Deuterium-burning
threshold of $\approx 13$ $M_J$ (where $M_J$ is the mass of Jupiter;
~\citealp{oppen00}) are defined as planets and included in the catalog,
with the exception of
companions with masses as large as 20 $M_J$ in systems already containing
one planetary mass companion.} (Jupiters on few-days orbits, very high eccentricities,
objects with masses five to ten times the mass of Jupiter)
had been predicted {\it a priori} by theoretical models. Correlations
among planetary orbital and physical parameters had not been anticipated.
The dependence of the frequency and properties of planetary systems on some
of the characteristics of the parent stars (mass, metallicity)
had not been foreseen.

The unexpected properties of extrasolar planets have sparked new
enthusiasm among theorists, who have engaged in fruitful
intellectual confrontations, with the aim to move from a set of
models describing separately different aspects of the physics of
the formation and evolution of planetary systems to a plausible,
unified theory capable of making robust and testable predictions.
By analogy, a number of new as well as old techniques of astronomy
has been fuelled by the new discoveries, with the twin goals to
follow-up and better characterize the extrasolar planet sample and
cover new areas of the discovery space. The result is an ongoing,
positive, creative tension between theory and observation that
will put to the test the most basic ideas of how planets form and
evolve.

Among the detection techniques, astrometry is the oldest. Nonetheless,
no planet discovery announcement has been ascribed to it
yet, and a few confirmations of previously detected systems have
made the news well after other novel methods, considered
completely unrealistic a few years back, had already achieved
important results.
However, the contribution of astrometric
measurements of sufficient precision is potentially very relevant
for a continued improvement of our understanding of the formation
and evolution of planetary systems, and possibly for the
identification of the first Earth-sized objects worthy of follow-up
observations to search for signs of the presence of essential
`bio-markers', i.e. of the existence of life as we know it outside
of our Solar System.

It is the aim of this review paper to
describe the current and future sensitivity of astrometric
measurements, both from the ground and in space, and to delineate the
areas of planetary science in which astrometry will
be able to make significant contributions, by comparison with
other direct and indirect methods for the detection and
characterization of planetary systems.
In Section 2 I summarize the observed properties of extrasolar
planets, in connection with renewed
theoretical efforts in the fields of planet formation and
evolution. I describe in Section 3 the astrometric methods and
instrumentation utilized to hunt for planets, in light of a number of
technological as well as astrophysical challenges to be faced
in order to achieve the required degree of measurement precision.
Section 4 contains a brief review of past and present efforts to detect
planets with astrometry. In Section 5 I discuss the planet finding
capabilities of future astrometric observatories.
Finally, in Section 6 I outline a number of experiments
that can be conducted with high-precision astrometry,
to illustrate its potential for important contributions to planetary science.

\section{Emerging Statistical Properties of Planetary Systems}

Ten years after the announcement of the first Jupiter-sized object orbiting
a star other than the Sun (\citealp{mayor95}), the number of
extrasolar planets announced has increased by a few tens
per year. Based on the large datasets mostly made available by ground-based 
Doppler surveys (e.g, \citealp{marcy04b}, and references therein; 
\citealp{mayor04}, and references therein), estimates of the 
frequency $f_p$ of giant planets around solar-type
(late F, G, and early K spectral types) stars in the solar neighborhood 
($D\lesssim 50$ pc) are all in fair agreement with each other
(\citealp{zucker01a}; \citeauthor{tabachnik02} 
\citeyear{tabachnik02}; \citeauthor{lineweaver03}
\citeyear{lineweaver03}; \citealp{marcy04b}; \citealp{naef04}).
Quoted values range between $f_p\simeq 4\%-5\%$ and $f_p\simeq 6\%-9\%$ for
planet masses in the range $1 M_J\leq M_p\leq 10 M_J$, and orbital radii 
$a\leq 3$ AU, and $0.5 M_J\leq M_p\leq 10 M_J$, and $a\leq 4$ AU, respectively.

On the observational side, the startling diversity of
planetary systems, when compared
with the properties of our own Solar System, has, if possible,
become more evident with the addition of newly discovered planets.
On the theoretical side, the only concept that has not
yet undergone significant revision or criticism is the paradigmatic statement
that planets form within gaseous disks around young T Tauri stars. Many old
ideas have been revisited or revived, and a number of completely new ones has
been proposed in an attempt to confront and explain the observational
data on extrasolar planets.

\subsection{Mass, Period, and Eccentricity Distributions}

The extrasolar planet sample exhibits many interesting and surprising
characteristics. The distribution of minimum masses
\footnote{Only the product $M_p\sin i$ between the actual companion mass and
the unknown inclination of the orbital plane with respect to the line of sight
can be derived from Doppler-shift measurements.} rises towards lower masses
(e.g., \citeauthor{tabachnik02} \citeyear{tabachnik02}, and references therein;
\citeauthor{lineweaver03} \citeyear{lineweaver03};
\citeauthor{marcy04a} \citeyear{marcy04a}). Incompleteness below
$M_p\sin i\sim 0.5$ $M_J$ becomes increasingly important, although recently
objects with minimum masses as low as the mass of Neptune have been discovered
(\citeauthor{butler04} \citeyear{butler04}; \citealp{mcarthur04};
\citealp{santos04b}). The sharp cut-off
at the high-mass tail of the distribution, above $M_p\sin i\sim 10-15$ $M_J$,
with very few low-mass companions to solar-type stars in the range
$10M_J \lesssim M_p\sin i \lesssim 80M_J$, 
and for orbital periods up to a decade, is often referred to as the 
``brown dwarf desert''.

Theoretical predictions of the mass distribution of extrasolar planets
within the context of the core-accretion model
(e.g., \citealp{lissa93}; \citealp{pollack96}) of giant planet formation
have recently been proposed (\citeauthor{alibert05} \citeyear{alibert05};
\citeauthor{ida04a} \citeyear{ida04a}, \citeyear{ida05}),
which qualitatively reproduce the observed one (particularly for objects
with $M_p\lesssim 5$ $M_J$). 
Rice et al. (\citeyear{rice03b}) and Rafikov (\citeyear{rafikov05})
argue instead that giant planets
formed by disk instability (e.g., \citealp{boss97}, \citeyear{boss01},
\citeyear{boss04}; \citealp{mayer02})
should preferentially populate the high-mass tail
($M_p\gtrsim 5$ $M_J$) of the planet mass distribution, while
Mayer et al. (\citeyear{mayer04}) indicate the possibility that the range of masses
of giant planets formed via core accretion and disk instability could
significantly overlap.

The period ($P$) and eccentricity ($e$) distributions also contain interesting 
features (\citeauthor{tabachnik02} \citeyear{tabachnik02}, 
and references therein; \citeauthor{udry03} \citeyear{udry03}; 
\citeauthor{marcy04a} \citeyear{marcy04a}; \citealp{halbwachs05}). 
Orbital periods are found in the range $1.5\lesssim P\lesssim 5400$ days.
About 20\% of the planet sample, the so-called ''hot'' Jupiters, are 
found orbiting within 0.1 AU. The number of planets increases with orbital
period for $60\lesssim P\lesssim 2000$ days, with increasing incompleteness
for orbital radii $\gtrsim 3$ AU. The median of the eccentricity
distribution of extrasolar planets is $\sim 0.3$. The orbits of
extrasolar planets span the whole range of available eccentricities, and
they can be extremely elongated (\citealp{naef01}).
Planets orbiting within 0.1 AU are all found with $e\simeq 0.0$,
a feature usually explained in terms of tidal circularization
(\citealp{goldreich66}).

Predictions on the actual orbital distance distribution of giant
planets have been made within the context of core-accretion models
which include mechanisms of inward orbital migration due to
tidal interactions between a gaseous disk and an embedded planet
(e.g., \citealp{goldreich79}, \citeyear{goldreich80}; \citealp{lin93};
\citealp{ward86}, \citeyear{ward97}). Trilling et al. (\citeyear{trilling02})
and Armitage et al. (\citeyear{armitage02}) were able to
qualitatively reproduce the observed semi-major axis
distribution of giant planets, for $a > 0.1$ AU. Similar results
were obtained recently by Alibert et al. (\citeyear{alibert04},
\citeyear{alibert05}) and by Ida \& Lin (\citeyear{ida04a},
\citeyear{ida05}). However, these models largely neglect
the difficult problem of identifying general mechanisms capable of
stopping orbital migration (e.g., \citealp{terquem03}, and references therein).
In the context of the disk instability model of giant planet formation,
migration efficiency might not be very effective (\citealp{rice03a},
\citeyear{rice03b}; \citeauthor{mayer04} \citeyear{mayer04}),
thus planets formed by this mechanism should be
found on not too close-in orbits (\citeauthor{rice03b} \citeyear{rice03b};
\citeauthor{rafikov05} \citeyear{rafikov05}).

The large spread of orbital eccentricities is difficult to explain
by the standard core-accretion model. Several
mechanisms have been proposed to reproduce the observed $e$ distribution,
which are based on dynamical interactions of various nature, such 
as interactions between the planet and a gaseous or planetesimal disk,
planet-planet resonant interactions, close encounters between planets,
or secular interactions with a distant companion star (for a review
see \citealp{tremaine04}, and references therein), but none of
them can represent alone the observed distribution in a natural way. Furthermore,
in multiple-planet systems different eccentricity excitation mechanisms induce
different evolution of the orbital alignment, and planetary orbits could be
significantly non-coplanar. The
alternative mode of planet formation by disk instability gives rise
to eccentric orbits, but no clear prediction of the final distribution
of eccentricities has been provided yet (e.g., \citealp{papalo01};
\citealp{terq02}; \citeauthor{mayer04} \citeyear{mayer04}).

\subsection{Correlations}

With improved statistics, in recent years a number of studies
has been carried out to find evidence of correlations among
orbital parameters and masses, and between
planet characteristics and stellar host properties.

As initially pointed out by Zucker \& Mazeh (\citeyear{zucker02}),
Udry et al. (\citeyear{udry03}) and more recently by Eggenberger et al.
(\citeyear{eggenberger04}), the extrasolar planet
sample exhibits a statistically significant lack of massive, close-in planets. These
objects are the easiest to detect with the Doppler method
\footnote{Recall that the radial-velocity amplitude $K\propto P^{-1/3}M_p\sin i$},
thus the paucity of high-mass planets on short-period orbits is
real, and not due to selection effects.

Regardless of the formation mode, orbital migration effects are the likely responsible for the observed
$M_p\sin i - P$ correlation. Many models can reproduce such results,
including reduced migration efficiency due to gap opening
(\citeauthor{ward97} \citeyear{ward97}; \citeauthor{trilling02}
\citeyear{trilling02}), substantial mass-loss through Roche lobe
overflow (\citealp{trilling98}; \citealp{gu03}), and accelerated
orbital decay due to enhanced tidal interactions with the host stars
(\citealp{patzold02}). Finally, Ida \& Lin (\citeyear{ida04a}) have derived a
theoretical mass-period diagram that closely resembles the
one of the extrasolar planet sample, and predicted a
paucity of planets in the intermediate mass range
$0.05\leq M_p\leq 0.5$ $M_J$, for orbital distances $< 3$ AU.

The possibility that super-solar metallicity could correspond
to a higher likelihood of a given star to harbor a planet has been
the subject of a large number of studies (for a detailed review
see \citealp{gonzalez03}). 
Recent works (e.g., \citeauthor{santos01} \citeyear{santos01},
\citeyear{santos04a}; \citealp{valenti05}) have conclusively shown that
planet occurrence correlates strongly with the host stars' primordial
metallicity. Up to $\sim 20\%$ of metal-rich ([Fe/H] $\gtrsim 0.3$) F-G-K
stars harbor planets, while less than 3\% of metal-poor stars
([Fe/H] $\lesssim 0.0$) have been found to be planet hosts.

Based on the core accretion model,
Kornet et al. (\citeyear{kornet05}) and Ida \& Lin (\citeyear{ida04b}) have quantified
the dependence of planetary frequency on stellar metallicity,
in qualitatively good agreement with the observed trend. The alternative
scenario of giant planet formation via disk instability, however,
is mostly insensitive to the primordial metal content of the
protoplanetary disk (\citealp{boss02}; \citeauthor{rice03b} \citeyear{rice03b}),
thus planet occurrence should not be hampered around metal-poor stars. 
The observed trend suggests that giant planet formation by core accretion 
predominates in the metal-rich regime ([Fe/H]$\simeq 0.0$). 
However, due to the low numbers of metal-poor stars ([Fe/H]$\lesssim -0.5$) 
surveyed to-date, no definitive conclusion can be drawn, except that maybe 
both mechanisms operate (\citealp{beer04}).

Several authors have searched in the past for possible correlations
between stellar metallicity and planet properties. Udry et al. (\citeyear{udry02}),
Santos et al. (\citeyear{santos01}, \citeyear{santos03}), and Fischer et al.
(\citeyear{fischer02}) searched for
correlations in the $M_p\sin i$-[Fe/H] and $e$-[Fe/H] diagrams, but concluded no
statistically significant trend can be found.
The $P$-[Fe/H] diagram deserves instead more attention. Gonzalez
(\citeyear{gonzalez98}) and Queloz et al. (\citeyear{queloz00})
initially argued that metal-rich stars
seem to possess an excess of very short-period planets with
respect to other planet hosts. In later works (\citeauthor{santos01}
\citeyear{santos01}, \citeyear{santos03}; \citealp{laws03})
no trend was found. However,
after removing some potential sources of bias, Sozzetti
(\citeyear{sozz04}) has shown how this trend is still present in the data,
specifically when one restricts the analysis to single planets
orbiting single stars.

If real, the $P-$[Fe/H] correlation could either reflect the fact that migration
rates are slowed down in metal-poor protoplanetary disks
(\citealp{livio03}), or it might be indicative of longer
timescales for giant planet formation around metal-poor stars, and thus
reduced migration efficiency before the disk dissipates
(\citeauthor{ida04a} \citeyear{ida04a}).

Finally, planet frequency appears to correlate with the primary
mass. In particular, as pointed out by e.g. Butler et al. (\citeyear{butler04}),
the occurrence rate of giant planets orbiting within 2 AU
around M dwarfs ($0.3 M_\odot$ $\lesssim M_\star\lesssim 0.6$ $M_\odot$) seems
suppressed by about an order of magnitude with respect to that
of analogous planets around F- and G-type dwarfs
($0.8 M_\odot$ $\lesssim M_\star\lesssim 1.3$ $M_\odot$).

The presently small number of giant planets discovered around
M dwarfs might still be partly an artifact due to the small-number statistics
(\citeauthor{butler04} \citeyear{butler04}). 
However, the observed trend is supported by theoretical arguments
(\citealp{laughlin04}; \citeauthor{ida05} \citeyear{ida05}) for a strong dependence
of planet occurrence rates on stellar mass, highlighted by a
significantly suppressed probability of forming giant planets
by core accretion around M dwarfs, and by an enhanced likelihood for
M dwarfs to harbor Neptune-sized objects. Planet occurrence, on the
other hand, may not be a strong function of the primary mass for
objects formed by disk instability (e.g., \citealp{boss00}).

\subsection{Multiple Systems and Planets in Stellar Systems}

Some 10\% of the planet hosts are found to be orbited by multiple systems,
containing up to 4 planets, while $\sim 12\%$ of the planet-bearing
stars are themselves components of wide multiple stellar
systems, and in two of the latter cases the stars harbor multiple-planet
systems.

A few authors have searched for differences between the distributions
of orbital elements and masses of planets orbiting single and multiple
stars and between those of single- and multiple-planet systems.
Zucker \& Mazeh (\citeyear{zucker02}) and Eggenberger et al.
(\citeyear{eggenberger04}) presented evidence for
no correlation between masses and periods of planets found in
stellar systems, while Marcy et al. (\citeyear{marcy04a}) compared
visually the eccentricity and mass distributions of single planets and
planetary systems, and concluded that no significant difference was
apparent. Finally, Mazeh \& Zucker (\citeyear{mazeh03})
have recently presented arguments for
a correlation between mass ratios and period ratios among pairs of planets
in multiple systems (assuming coplanarity of the orbits).

From a theoretical viewpoint, the overall impact of the presence of a
secondary star on the efficiency of planet formation and migration
is far from being clear. For example, Nelson (\citeyear{nelson00}) and
Mayer et al. (\citeyear{mayer05}) argue that giant planet formation by either core
accretion or disk instability can be strongly inhibited in
relatively close binary systems with separation of order of a few tens
of AUs. Boss (\citeyear{boss98}), however, comes to opposite conclusions. Due to
enhanced migration and gas accretion rates (\citealp{kleybur00};
\citealp{kley00}, \citeyear{kley01}; \citealp{nelson03})
planets formed around binaries
should not show evidence for a mass-period correlation. These
predictions appear to agree with the observed trend.

Theoretical investigations of the long-term dynamical evolution of
multiple-planet systems (e.g., \citealp{kiseleva02}; \citealp{ji03}, 
and references therein; \citealp{correia04}; \citeauthor{barnes04} 
\citeyear{barnes04}; \citeauthor{godz04} \citeyear{godz04}, and references therein) 
have allowed to divide such systems in three broad classes:
$a)$ ``hierarchical'' planetary systems, with widely separated orbits,
in which dynamical interactions appear negligible;
$b)$ planetary systems subject to strong secular interactions;
$c)$ planetary systems locked in mean motion resonances,
which in some cases exhibit important variations of the orbital elements
on time-scales comparable to the time-span of the radial-velocity monitoring. 
In multiple-planet systems, regions of dynamical stability do exist inside
the parent stars' Habitable Zones
\footnote{The Habitable Zone of any star is defined as the range
of orbital distances at which a potential water reservoir, the primary
ingredient for the development of life as we know it, would be
found in liquid form (e.g., \citealp{kasting93})}, where Earth-sized
planets may be found (e.g., \citealp{menou03}, and references therein;
\citealp{jones05}, and references therein). 
Furthermore, the detected giant planets in binaries are likely to
reside in stable orbital configurations, and there are margins
for the presence of rocky planets in the Habitable Zone of
close binaries (e.g., \citealp{holman99}; \citealp{pilat03}, and
references therein; \citealp{marzari05}; \citealp{musielak05},
and references therein), although the formation of terrestrial planets
in such environments does not easily occur in the first place
(\citealp{thebault02}; \citealp{raymond04};
\citealp{barnes04a}; \citealp{raymond04a}). However, the lack of information
on the actual mutual inclination angles between pairs of planetary orbits
somewhat limits the generality of these findings.

\subsection{Planetary Radii and Atmospheres}

A handful of hot Jupiters ($P\lesssim 4$ days) have been discovered
by means of photometric transit surveys (\citeauthor{uda02a} \citeyear{uda02a},
\citeyear{uda02b}, \citeyear{uda03}; \citealp{alonso04}),
and confirmed by high-resolution
spectroscopic measurements (\citealp{torres04a}; \citealp{bouchy04};
\citealp{moutou04}; \citeauthor{pont04} \citeyear{pont04}; 
\citeauthor{sozzetti04} \citeyear{sozzetti04}; 
\citealp{konacki03}, \citeyear{konacki04}, \citeyear{konacki05}),
while one, HD 209458b, was observed transiting (\citeauthor{charbon00} 
\citeyear{charbon00}; 
\citealp{brown01}) subsequently to the detection of its gravitational
pull on the star (\citeauthor{mazeh00} \citeyear{mazeh00}; \citealp{henry00}).

The combination of the Doppler-shift and transit photometry measurements
allows to derive estimates of the true mass and radius of the planet.
These two critically interesting
parameters can then be used for directly constraining structural
models of irradiated giant planets (see \citealp{guillot05}, and references
therein, for a detailed review). 
The measured radii for six of the seven transiting planets provide
good agreement with theoretical expectations, while all models
seem to systematically underestimate the radius of HD 209458b. The
recent successful detection of thermal emission in the infrared
from the planet (\citealp{deming05}), and in particular the timing
of the secondary eclipse, clearly suggests that the planet revolves on
an orbit with no significant eccentricity, essentially ruling out
mechanisms invoked to provide additional heat/power sources in the
core, such as tidal dissipation of a nonzero eccentricity induced
by the gravitational perturbation of an undetected long-period
companion (\citealp{bode01}, \citeyear{bode03};
\citealp{laughlin05}).

Transmission spectroscopy during transits has allowed to detect
absorption features in the spectrum of HD 209458 which are
indicative of the presence of various constituents in the planet's
atmosphere, notably sodium, hydrogen, oxygen, and carbon
(\citealp{charbon02}; \citeauthor{vidal03} \citeyear{vidal03}, 
\citeyear{vidal04}). Also, the planet appears to have an extended atmosphere,
presumably due to evaporation effects. In two cases, detection of
the planet's thermal emission (\citealp{charbonneau05};
\citealp{deming05}) has permitted to estimate the planets'
effective temperatures and Bond albedos, and to infer the presence
of atmospheric water vapor and carbon monoxide.

Theoretical predictions of the atmospheric composition, temperature, and
circulation of irradiated giant planets (\citealp{burrows05a};
\citealp{fortney05}. For a review see \citealp{burrows05}) 
are in fair agreement with the first infrared direct detections,
however a proper understanding of the fine details of the emergent
spectra of TrES-1 and HD 209458b will require both improved
modelling and larger, high-quality datasets. Finally, studies of
the phenomenon of atmospheric escape from hot Jupiters
(\citealp{lammer03}; \citealp{gu03}, \citeyear{gu04};
\citealp{lecavelier04}; \citealp{baraffe04};
\citealp{griessmeier04}) predict that, under strong irradiation,
these objects, depending on their mass and orbital distance, could
undergo significant evaporation of their gaseous envelope, in
reasonable agreement with observations.

\subsection{Toward a Unified Picture}

The observational data on extrasolar planets show such striking
properties that one must infer that planet formation and evolution
is a very complex process. Indeed, the confrontation between
theory and observations indicates that there are numerous problems in connection
with the elucidation of planetary formation and evolution processes.

An ideal theory of planet formation and evolution
should be capable of explaining in a self-consistent way, be it
deterministic or probabilistic, all the different properties of
planetary systems discussed
above. To this end, the help from future data obtained with a
variety of different techniques will be crucial. Ultimately,
both theory and observation will have to provide answers to
a number of fundamental questions, that can be summarized as
follows. (1) Where are the earth-like planets, and what is
their frequency?
(2) What is the preferred method of gas giant planet formation?
(3) Under which conditions does migration occur and stop?
(4) What is the origin of the large planetary eccentricities?
(5) Are multiple-planet orbits coplanar?
(6) How many families of planetary systems can be identified
from a dynamical viewpoint?
(7) What are the atmospheres, inner structure and evolutionary
properties of gas giant planets?
(8) Do stars with circumstellar dust disks actually shelter
planets?
(9) What are the actual mass and orbital elements
distributions of planetary systems?
(10) How do planet properties and frequencies depend upon
the characteristics of the parent stars (spectral type, age,
metallicity, binarity/multiplicity)?

With the above questions in mind,
I focus next on what the contribution of astrometry from ground
and in space will be, by presenting a summary of methods and instrumentation,
by reviewing past and present efforts, by discussing future prospects
and by putting this technique in perspective with other planet-detection
methods.

\section{Astrometric Planet Detection Techniques}

Astrometric detection of extrasolar planets can be conducted with instrumentation
on the ground or in space. I describe in this Section the
generic approach to planet detection and measurement with astrometry,
in terms of what type of data are needed, how to extract and model
the planet signal from the data in presence of a number of noise sources,
and how to assess the significance of a detection. I will conclude
the Section by summarizing results from a set of ground-based and
space-borne experiments aimed at demonstrating the theoretical predictions
on the achievable astrometric precision under a variety of conditions.

The general analysis methods can be applied to astrometric
observables appropriately defined for both monopupil and
diluted-aperture telescopes, both from the ground and in space.
To this end, I will describe the techniques in terms of the basic
observable and noise models and the estimation process. The observable model
produces theoretical values for the data as a function of adjustable
parameters. The noise model describes errors that corrupt the
astrometric data. The estimation process finds parameter values that
produce the closest agreement between the observable model estimates
and the data in light of the noise model.

\subsection{Observable Model}

The astrometric observable is generally defined as the angular position of
a star as measured by a given instrument in its local frame of
reference. The measurement could be
for example intrinsically one-dimensional, as is the case for space missions
such as ESA's $Hipparcos$ (\citealp{perryman97}) and $Gaia$ (\citealp{perryman01}),
which are designed to perform angular position measurements in their sensitive
directions by centroiding their diffraction-limited images. Or,
it could be a set of two coordinates on the focal plane of the instrument,
as is the case for ground-based telescopes (e.g, \citeauthor{gatewood87}
\citeyear{gatewood87};
\citeauthor{dekany94} \citeyear{dekany94}; \citeauthor{pravdo96}
\citeyear{pravdo96}). Finally, it could
be either the optical path-length difference between the two arms of an
interferometer on the ground (\citealp{shao88}; \citealp{armstrong98};
\citealp{colavita99a}; \citealp{vanbelle98}; \citealp{glindemann03})
or in space, such as NASA's Space Interferometry Mission
($SIM$; \citealp{danner99}), or the normalized difference between the signals
of two photomultiplier tubes (the Transfer Function) of a space-borne
interferometer, such as $HST$/FGS (\citealp{taff90}).

Both from the ground and in space, astrometric measurements can be
performed in wide-angle mode, i.e relative to a local frame of reference composed
of a set of one or more reference stars at typical angular distances of
several degrees from the target object.
If the selected local reference frame lies at $\lesssim 1^\circ$, the data
are said to be collected while operating in narrow-angle mode. From space, without
the limiting presence of atmospheric turbulence, which induces large-scale
wavefront distortions (\citealp{lindegren80}), the astrometric observable can be
determined with respect to a global inertial reference frame by accurately
bridging together multi-epoch observations of objects distributed everywhere
in the sky (and thus separated by typically tens of degrees) and by adopting
a global closure condition over the whole celestial sphere.
The combination of such an observing scenario and data reduction
method is called global astrometric mode (\citealp{kovalevsky80}).

Regardless of the mode of operation and of the instrument utilized to
carry out the measurements, four categories of information should be identified
for inclusion in the observable model utilized to calculate theoretical values
of the observable with negligible errors:
(1) the location and motion of the target (if working in global astrometric
mode) and a possible set of reference stars (if working in wide-angle or
narrow-angle mode), (2) the location and motion of the observing instrument
(if on the ground) or the attitude of the spacecraft (if in space),
(3) the number, masses, and orbital parameters of companions to the
target (and reference stars where applicable), and (4) any physical effects
that modify the apparent positions of stars.

\subsubsection{Stellar and Instrumental Parameters}

The star information consists of the five basic astrometric
parameters--position on the celestial sphere (2
parameters, say $\lambda$ and $\beta$),
proper motion (2 parameters, say $\mu_\lambda$ and $\mu_\beta$),
and parallax (1 parameter, say $\pi$)--plus the radial velocity $v_r$,
which can be determined by auxiliary measurements or
from a sufficiently large secular acceleration.

The stellar locations and motions are usually determined
in the Solar-System barycentric frame in which the global frame
is defined. A variety of transformations can be utilized to connect
the instrument-specific observational frame to the stellar frame.

If the object's position in the instrument and barycentric frame are
described by the vectors $\mathbf{Z}$ and $\mathbf{S_\star}$, respectively,
then for an all-sky survey instrument such as $Hipparcos$ or $Gaia$
the mapping is specified by the $3\times 3$ rotation matrix $\mathsf{A}$:
\begin{equation}
\mathbf{Z} = \mathsf{A}\mathbf{S_\star}
\end{equation}
From this relation, the along-scan angular coordinate of the object,
which constitutes the actual observable, can be solved for in
terms of $\mathsf{A}$ and $\mathbf{S_\star}$.
The matrix $\mathsf{A}$ is a continuous function of time that
specifies the spacecraft attitude. It could be defined by a
set of nine functions $A_{ij}(t)$, $i,j = 1,2,3$. Or, it could
be expressed in terms of three Euler angles ($\phi(t)$, $\theta(t)$,
$\psi(t)$), as it was done for $Hipparcos$. Alternatively, it could
be described by means of the quaternion representation, as is
presently envisioned for $Gaia$.

For an interferometer operating in wide-angle mode, both
on the ground and in space, the measured optical path-length delay
$d_\star$ corresponds to the instantaneous three-dimensional position of
the target on the sky projected onto the interferometer baseline:
\begin{equation}
d_\star = \mathbf{B}\cdot\mathbf{S_\star} + C
\end{equation}
The baseline vector $\mathbf{B} = B \mathbf{u_b}$ of
length $B$ describes the spacecraft attitude, while $C$ is a constant
term representing residual internal optical path differences.
In the narrow-angle regime, this expression is modified as follows:
\begin{equation}
\Delta d_{\star,j} = \mathbf{B}\cdot(\mathbf{S_\star}-\mathbf{S_j})
\end{equation}
The relative optical path-length delay $\Delta d_{\star,j}$ is
then the instantaneous angular distance between the target and
its $j$th reference star projected onto $\mathbf{B}$. Due to the
differential nature of the measurement, the constant term $C$ cancels
out, to first order.

Finally, in order to relate the detector frame of a ground-based
monolithic telescope or $HST$/FGS to the actual coordinates of
an object in the sky, a plate-reduction transformation is applied
(e.g., \citealp{kovalevsky04}). In this case, the two-dimensional
standard cartesian coordinate vector $\mathbf{s}(s_1,s_2)$
describing the position of the target in the plane of the sky
is mapped into the two-dimensional vector of measured coordinates
on the detector $\mathbf{r}(r_1,r_2)$ via the transformation:
\begin{equation}
\mathbf{r} = \mathsf{M}\mathbf{p}
\end{equation}
In the above expression, $\mathsf{M}$ is the model matrix:
\begin{equation}
\mathsf{M} = \pmatrix{s_1 & s_2 & 1 & 0 & 0 & 0 & \dots\cr
     0 & 0 & 0 & s_1 & s_2 & 1 & \dots\cr}
\end{equation}
The column vector $\mathbf{p} = (p_1,p_2,\dots,p_n)$ contains
the so-called {\it plate constants}. A minimum of six is required,
to describe scale and rotation factors and offsets of coordinate
origins between the two frames of reference, but it is not uncommon
to include focal-plane tilt and other optical distortion terms,
in addition to terms dependent on the magnitude and color index
of the star observed. The same relation holds for all the objects
used as reference stars.

\subsubsection{Planet Parameters}

Masses and orbits of companions to the target object (and
reference stars where applicable) come from fitting a model of Keplerian
orbital motion to the data. The Keplerian orbit of each companion is
described by seven parameters: semi-major axis $a$ with respect to
the center of mass of the system,
period $P$, eccentricity $e$, inclination $i$, position angle
of the line of nodes $\Omega$, argument of pericenter $\omega$,
and epoch of pericenter passage $\tau$.

The observable model computes the star's reflex motion projected on the
plane of the sky due to the gravitational pull of such companions, that
might be stellar or sub-stellar (brown dwarfs and planets) in nature.
If the primary mass is $M_\star$ and the secondary is a planet of
mass $M_p$, then, assuming a perfectly circular orbit, the apparent
amplitude of the perturbation, i.e. the stellar orbital radius around
the center of mass of the system scaled by the distance from the observer,
is the so-called \textit{astrometric signature}:
\begin{equation}  \label{signature}
\alpha = \frac{M_\mathrm{p}}{M_\star}\frac{a}{D}
\end{equation}
If $M_\mathrm{p}$ and $M_\star$ are given in solar mass units,
$a$ in AU, and $D$ in parsec, then $\alpha$ is in arcsec.

Table~\ref{alpha} summarizes the values of $\alpha$ for a range of planet
masses at different orbital radii from a 1-$M_\odot$ star at 10 pc, compared
to typical values of parallax and proper motion for stars in the solar
neighborhood.

As one can see, planetary signatures are a higher-order effect for
astrometry. Jupiter-sized
planets already produce perturbative effects whose size is smaller
than the typical $Hipparcos$
milli-arcsecond (mas) measurement precision. Detection of orbital motion
induced by terrestrial planets necessarily implies an improvement of
2-3 orders of magnitude in precision, down to the few micro-arcseond ($\mu$as)
regime.

Finally, in the case for example of a multiple-planet system,
simply considering independent Keplerian orbits
might not be sufficient, whenever secular or resonant gravitational
perturbations among
planets in the systems (due to the presence of large mass
ratios, highly eccentric orbits, commensurabilities between orbital
periods, and significantly non-coplanar orbits) are strong enough to induce
measurable variations of orbital elements over time-scales comparable to
the time-span of observations. For these cases,
additional information must be fed to the observable model,
such as approximate analytical expressions describing the gravitational
perturbations and consequent time variations of the orbital elements,
or fully self-consistent fitting algorithms which include the direct
solution of the equations of motion of an N-body system.

\subsubsection{Physical Effects}

A variety of physical effects that cause the apparent coordinates
of observed stars to differ from the transformed
values of their true barycentric coordinates can be taken into
account in principle. In order to understand which effects are
more relevant, the driver is the limiting single-measurement 
precision the adopted instrument is designed to achieve.
The 1 mas state-of-the-art astrometric precision has been set by
$Hipparcos$ and $HST$/FGS. The expected improvement in measurement
precision by a few orders of magnitude envisioned for future ground-based
and space-borne instrumentation such as VLTI, Keck-I, $SIM$, and $Gaia$
will sensibly increase the order of higher approximations.

Higher-order perturbations can be classical in nature, such as
additional secular variations in the target space motion with respect to
the observer, or intrinsically relativistic, such as corrections to classical
effects due to the motion of the observer itself, or contributions
coming from the gravitational fields of massive bodies in the
vicinity of the observer. Many of these effects are well-known in pulsar
timing work, and are included in detailed models of pulse arrival
times (e.g., \citealp{hellings86}; \citealp{wols92}; \citealp{stairs98},
\citeyear{stairs02},
and references therein). However, they are often neglected in astrometric
data reduction with mas-level precision.

Secular changes in proper motion (perspective acceleration)
and annual parallax can be quantified as time-derivatives
of these two astrometric parameters (e.g., \citealp{dravins99},
and references therein):

\begin{eqnarray}
\frac{\mathrm{d}\mu}{\mathrm{d}t}& = & -\frac{2v_r}{AU}\mu\pi\\
\frac{\mathrm{d}\pi}{\mathrm{d}t}& = & -\frac{v_r}{AU}\pi^2
\end{eqnarray}

The above quantities are expressed in arcsec yr$^{-2}$ and arcsec yr$^{-1}$,
respectively, if the radial velocity $v_r$ is in km s$^{-1}$, $\pi$ is in arcsec,
$\mu$ is in arcsec yr$^{-1}$, and the astronomical unit
$AU = 9.77792\times 10^5$ arcsec km yr s$^{-1}$.

I show in Figure~\ref{changepimu} the values of $\mathrm{d}\mu/\mathrm{d}t$
and $\mathrm{d}\pi/\mathrm{d}t$ as a function of $v_r$ and of the product
$v_t\times v_r$, where the tangential velocity $v_t = AU\mu/\pi$
(with the astronomical unit now defined as $AU = 4.74$ km yr s$^{-1}$).
As one can see, the effect of
changing annual parallax is below a few $\mu$as/yr for stars
more distant than a few pc from the Sun, even assuming large
values of $v_r$. Its inclusion in the observable model might then
be limited to the nearest stars.
The contribution from perspective acceleration drops
below a few $\mu$as/yr$^2$ only for objects farther away than a few
tens of pc, thus such corrective term might have to be taken into
account in the observable model for a few hundred
nearby, high velocity stars.

Relativistic corrections to the motion of the observer with respect
to the solar-system barycenter (aberration) can be quantified
through the formula (e.g., \citealp{klioner03}; \citealp{kovalevsky04}):

\begin{eqnarray}
\alpha_\mathrm{aberr} & \simeq & \frac{v}{c}\sin\vartheta
            -\frac{1}{4}\frac{v^2}{c^2}\sin 2\vartheta \nonumber\\
                      && +\frac{1}{6}\frac{v^3}{c^3}\sin \vartheta
             (1+2\sin^2\vartheta)\,\,+ O(c^{-4}),
\end{eqnarray}
where $\vartheta$ is the angular distance between the direction to the
target and the observer's space velocity vector, $v$ is the modulus of
the space velocity vector of the observer, and $c$ is the speed of light.

The magnitude of the classic aberration term (first order in $v/c$) is
$\sim 20-30$ arcsec, while the second-order relativistic correction
amounts to $1-3$ mas. For a targeted measurement precision of 1 $\mu$as,
terms of order $(v/c)^3$ must be retained, but may be dropped for less
stringent requirements. In addition, for relativistic stellar aberration to be
properly accounted for, the spacecraft's velocity will need to be
determined to an accuracy of 20 mm/sec or better.

The purely general relativistic effect of deflection of light by Solar-System
objects is instead defined by the expression (e.g., \citealp{klioner03}):

\begin{equation}
\alpha_\mathrm{defl} = \frac{(1+\gamma)GM}{R_0c^2}\cot\frac{\psi}{2},
\end{equation}
where $\psi$ is the angular distance between the given Solar-System
body and the source, $G$ is the gravitational constant, $R_0$ is the distance
between the observer and the Sun, $M$ is the mass of the perturbing body,
and $\gamma = 1$ is the Post-Post-Newtonian (PPN) parameter.

I summarize in Table~\ref{deflect} the magnitudes of the gravitational effects
on limb-grazing light rays induced by all Solar-System planets and some of
the largest moons. If the targeted measurement precision is of
order of 1-10 $\mu$as, the observable
model should consider deflection by all these bodies,
when observing close to their directions, and even quadrupole effects
induced by the non-oblateness of the major Solar-System 
planets may have to be considered (e.g., \citealp{klioner03}).

Indeed, several attempts have been made in the past years (\citealp{soffel89};
\citealp{brumberg91}; \citealp{klioner92}; \citealp{defelice98},
\citeyear{defelice01}, \citeyear{defelice04}; \citealp{vecchiato03};
\citealp{klioner03})
to develop schemes for the reduction of astrometric observations
at the $\mu$as precision level directly within the framework of
General Relativity, either employing non-perturbative approaches or
the PPN formulation (\citealp{will93}). If a complete
general relativistic observable model is implemented, other more
subtle effects arise, such as the need to re-define parallax,
proper motion, and radial velocity, depending on the given choice
of the local reference system of the observer (whose motion must also
be described in physically adequate relativistic terms, see for
example \citealp{klioner04}), and the fact that these
parameters, when higher-order terms are included, cannot be
considered anymore independent from each other. Furthermore,
a number of possible effects that may be caused by gravitational
fields produced outside of the Solar System might
have to be considered, such as weak gravitational lensing by
distant sources (e.g., \citealp{belokurov02}), binary systems
viewed edge-on which contain compact objects such as neutron stars
and/or black holes (\citealp{doroshenko95}), and gravitational
waves (\citealp{kopeikin99}).

\subsection{Noise Model}

Astrometric data contain correlated and uncorrelated instrumental,
atmospheric (if operating from the ground), and astrophysical
noise. The noise model describes these uncertainties for use in
the estimation process, which, particularly by taking proper
account of correlated errors, provides the most accurate and
sensitive results. When a noise source can be identified and
modelled deterministically, such as a newly-found companion to a
reference star, its predictable effects can be incorporated into
the observable model and any provision in the noise model is
deleted. When uncertainties can be quantified but not modelled,
they are accounted for in the noise model.

I summarize below a variety of known sources of astrometric noise.
Where applicable, i.e. in the case of instrumental and atmospheric
noise, the different implications for filled-aperture telescopes and
interferometers will be discussed separately. For instrumental noise,
a further distinction will be made depending on whether astrometric
measurements are performed from the ground or in space.
The relative importance of any given source of noise will be gauged
bearing in mind the goal of achieving a final astrometric precision
of order of a few $\mu$as.

\subsubsection{Instrumental Noise}

The various sources of instrumental uncertainties in astrometric
measurements can be described in terms of
the two general classes of random, photon errors $\sigma_\mathrm{ph}$
and systematic errors $\sigma_\mathrm{sys}$.

The expression for the photometric noise of a monopupil
telescope is (\citealp{lindegren78}):
\begin{equation}\label{sigphot}
\sigma_\mathrm{ph} = \frac{\lambda}{4\pi A}\frac{1}{SNR}
\end{equation}

Here $A$ is the telescope aperture and $\lambda$ the monochromatic
wavelength of the observations (both in meters), while
$SNR$ is the signal-to-noise ratio of the target,
including sky background and detector noise
($SNR\propto\sqrt{N}\propto\sqrt{t}$, where $N$ is the number of
photoelectrons and $t$ is the exposure time in seconds).
For a $m_v = 15$ solar-type star observed near the zenith with good
(better than 0.5 arcsec) seeing conditions, and assuming an overall
system efficiency $\varepsilon = 0.4$ (including CCD quantum efficiency,
atmospheric and optics transmission), the contribution from
$\sigma_\mathrm{ph}$ over small fields of view ($< 1$ arcmin) can be
$\sim 300$ $\mu$as, $\sim 30$ $\mu$as, and $\sim 3$ $\mu$as in 1 hr integration
for  $A = 1$ m, $A = 10$ m, $A = 100$ m, respectively (see for example
\citealp{allen00}).

For an interferometer, the photometric error is expressed as
(\citealp{shao92}):

\begin{equation}
\sigma_\mathrm{ph} = \frac{\lambda}{2\pi B}\sqrt{\frac{t_c}{t}}\frac{1}{SNR},
\end{equation}
with $A$ replaced by $B$ in Eq.~\ref{sigphot}. The
atmospheric coherence time $t_c$, with average seeing conditions,
is typically a few tens of ms in the near-infrared $K$ band
(\citealp{shao92}; \citeauthor{quirrenbach94} 
\citeyear{quirrenbach94}; \citealp{lane03}),
while the signal-to-noise ratio
per coherence time $SNR$ is a measure of the uncertainty $\sigma_\phi$
in the measurement of the phase of the interferometric fringes
($\sigma_\phi\approx (\mathrm{SNR})^{-1}$, see for example \citealp{wyant75}).
The value of $SNR$ in this case is not only a function of the
number of counts $N$, dark count and background, and read-noise
(as in the filled-aperture case), but also of the square of the complex visibility $V^2$
($SNR\propto\sqrt{NV^2}$. See for example 
\citeauthor{quirrenbach94} \citeyear{quirrenbach94}; \citealp{colavita99}).

If the interferometer sensitivity is limited by the actual
atmospheric coherence time, accurate measurements of the fringe
phase, and thus fringe tracking, can be performed only on
very bright targets (typically $m_k\lesssim 5.0$), for which enough
photons are collected in a coherence volume $t_c r^2_c$ (where $r_c$
is the coherence diameter).
When the interferometer is used in phase-referencing mode (e.g.,
\citealp{lane03}, and references therein),
$t_c$ can be artificially extended by more than an order of
magnitude, thus allowing to improve the limiting magnitude of
the instrument, or allowing for higher $SNR$ at the same magnitude.
The fundamental requirement is that target and reference object
be separated by less than an isoplanatic angle $\theta_i$
\footnote{The isoplanatic angle is the angle in the sky over which
atmosphere-induced motion is well-correlated}, usually of order of tens
of arcsecs (e.g., \citealp{shao92}).
The isoplanatic angle $\theta_i\propto r_c/h\propto \lambda^{6/5}$
(with $h$ the effective height of the turbulence profile. See for
example \citealp{colavita94}), thus, as chances of finding reference objects
are increased in larger fields, the obvious choice is to use the
instrument in the infrared.
It is however with long baselines that interferometers have a major
photon-noise advantage. For a $m_k = 13$ star, $B = 100$ m, $SNR = 5$,
$t_c = 100$ ms, $\sigma_\mathrm{ph}\simeq 1$ $\mu$as in 1 hr integration can be
achieved (e.g., \citealp{shao92}).

The instrumental systematic term for a monolithic telescope can be for
example (\citeauthor{pravdo96} \citeyear{pravdo96}) expressed as:

\begin{equation}
\sigma_\mathrm{sys} = \sqrt{\sigma^2_\mathrm{CCD}+
            \sigma^2_\mathrm{OP}}
\end{equation}
The systematic limitations imposed by the CCD detectors
(charge transfer efficiency, deviations from uniformity or from
linear pixel response, etc.) through $\sigma_\mathrm{CCD}$ and optics imperfections
(optical aberrations and distortions, pixelization, etc.) through
$\sigma_\mathrm{OP}$ can be overcome with improvements
in image detector and optics technology.
Centroid accuracies of 1/100 of a pixel are today readily
achievable (\citealp{monet92}), translating in a typical value of
$\sigma_\mathrm{CCD}\simeq 50$ $\mu$as. Future developments
promise improvements of about one order of magnitude in
CCD image location accuracy (\citealp{gai01}), with hopes
to keep $\sigma_\mathrm{CCD}\lesssim 5-10$ $\mu$as.
Optical aberrations and distortions for large apertures ($> 5$ m) and
small fields of view ($< 1$ arcmin) are typically small. Pravdo
\& Shaklan (\citeyear{pravdo96}) have shown how for the Keck 10-m telescope
$\sigma_\mathrm{OP}\simeq 5$ $\mu$as, or less.

For narrow-angle measurements with an interferometer,
the systematic term will read (e.g., \citealp{shao92}):

\begin{equation}
\sigma_\mathrm{sys} = \sqrt{\sigma^2_\mathrm{l}+
                \sigma^2_\mathrm{B}}
\end{equation}

The two main sources of systematic errors arise from the uncertainty
$\sigma_\mathrm{l} = \delta l/B$ with which optical delay lines in long-baseline
interferometers can control internal optical path delays, and
from the uncertainty on the knowledge of the interferometric
baseline $\sigma_\mathrm{B} = (\delta B/B)\vartheta$, where $\vartheta$
is the angular separation between the target and a reference
star. As for the former, in order to reach a positional measurement precision of
$10$ $\mu$as with $B = 200$ m (the maximum baseline of the VLTI),
measurements of optical paths must be made with an accuracy of
$\sigma_\mathrm{l} < 10$ nm, a challenging but not impossible
achievement with today's technology (see for example \citealp{shao88}).
Due to the differential nature of the measurement, instead, knowledge of
the instrument baseline does not need to be very precise. For $B = 200$ m,
$\vartheta\approx 20$ arcsec, the goal of 10 $\mu$as precision is achieved
by determining the baseline stability with an uncertainty
$\sigma_\mathrm{B}\simeq 50$ $\mu$m, a requirement relaxed by a few orders
of magnitude with respect to a wide-angle measurement (e.g., \citealp{shao90}).

Finally, if the astrometric measurements are performed in space,
additional random and systematic error sources must be taken into
account, which are introduced by the satellite operations and environment.
For example, a class of relevant error sources is related to the determination of
the spacecraft attitude. Attitude errors can occur due to perturbations
produced by the solar wind, micrometeorites, particle radiation,
and radiation pressure. Thermal drifts and spacecraft jitter can also
induce significant errors in attitude determination.
However, these noise sources are hard to quantify {\it a priori} in
a very general fashion. Detailed error models must be developed
case by case (see Section 4.5).
Thus, in the design of a space-borne observatory for high-precision astrometry,
be it a monolithic telescope or an interferometer, {\it ad hoc}
calibration procedures must be devised, in order to attain the
goal of limiting the magnitude of such instrumental error sources at
the few $\mu$as level.

\subsubsection{Atmospheric Noise}

For ground-based instrumentation, the atmosphere constitutes an
additional source of noise through both its turbulent layers
(a random component) and due to the differential chromatic refraction
(DCR) effect (a systematic component).

The problems caused by the DCR effect can be very difficult to overcome.
The magnitude of the effect depends on zenith distance $z_0$, air
temperature and pressure, spectral band and star color, and even the
non-sphericity of the Earth (e.g., \citealp{gubler98}).
The precision of theoretical and
empirical DCR correction formulae degrades very quickly with $z_0$.
Even at small zenith distances, for a monolithic telescope
the goal of $\mu$as astrometry is unlikely to be attained. Typical
uncertainties for small values of $z_0$ can be of order of
$\sigma_\mathrm{DCR}\approx 1-3$ mas, and increase by over an
order of magnitude close to the horizon (\citealp{kovalevsky04}).

For conventional narrow-angle astrometric measurements with
separations of $10-30$ arcmin, the positional
error $\sigma_\mathrm{atm}$ due to atmospheric turbulence
is weakly dependent on separation and does not depend on
$A$ (or $B$). This prevents the achievement of sub-mas
astrometric precision (e.g., \citealp{lindegren80}; \citeauthor{han89} 
\citeyear{han89}). For separations $< 1-10$ arcmin, the situation improves. 
In this regime, the astrometric error due to anisoplanatism for a
filled-aperture telescope has been calculated
in the past (\citealp{lindegren80}; \citealp{shao92}) as:
\begin{equation}\label{linde}
\sigma_\mathrm{atm} \approx 300 A^{-2/3}\vartheta t^{-1/2}
\end{equation}
In this case $\vartheta$ is in radians, $A$ is expressed in meters, and
the factor 300 is obtained directly from the phase-structure
function describing the turbulence, assuming standard Kolmogorov-Hufnagel
(\citealp{hufnagel74}) atmospheric and wind-speed profiles
for good seeing conditions (FWHM $\approx 0.5$ arcsec), typical of a site
such as the Keck Observatory (e.g., \citealp{shao92}).
The rather weak power dependency of $\sigma_\mathrm{atm}$ on
target-reference star separation and objective diameter implies
that a typical value $\sigma_\mathrm{atm}\simeq 1$ mas is obtained
with $A = 1$ m in $t = 1$ hr of integration with a separation $\vartheta = 1$
arcmin. In order to suppress $\sigma_\mathrm{atm}$ by two-three orders
of magnitude one would require unrealistically large $A$ and $t$, unless
$\vartheta$ is limited to uselessly small angles.

Interferometers can in principle get much closer to the limits
in precision set by the atmosphere. For a diluted architecture,
in fact, Eq.~\ref{linde} now reads
$\sigma_\mathrm{atm} \approx 300 B^{-2/3}\vartheta t^{-1/2}$.
Values of $B$ of order of 100-200 meters are more easily attainable
than the equivalent filled-aperture size. Thus, for a 20-arcsec
star separation and a 200-m baseline, a 1-hr integration would allow to
achieve $\sigma_\mathrm{atm} \simeq 10$ $\mu$as (\citealp{shao92}).
In conditions of extremely favorable seeing, large isoplanatic angles,
and long atmospheric coherence times, such as those reported above Dome C
in Antarctica (\citealp{lloyd02}; \citealp{lawrence04}), atmospheric
errors in image motion maybe reduced by another order of magnitude.

However, if phase-referencing is used to artificially increase $t_c$
and the limiting $m_k$, additional noise sources are introduced.
In particular, coherence losses occur due to
fluctuations in the fringe position during integration, which induce
in turn a visibility reduction by a factor
$\eta = \mathrm{e}^{-\sigma^2_{\Delta\phi}}$ (where $\sigma_{\Delta\phi}$
is a measure of jitter in the referenced phase. See for example
\citealp{colavita94}). This in turn limits the achievable $SNR$ in a given $t_c$,
thus contributing to increase $\sigma_\mathrm{ph}$. These time-dependent
effects can be divided
into two classes, namely instrument-specific errors in the
determination of the phase, and those that are due one more time
to atmospheric propagation effects. The dominant effect 
(\citeauthor{quirrenbach94} \citeyear{quirrenbach94}) is again induced by 
DCR. For large values of $z_0$, $\eta\longrightarrow 0$, thus
applications of this technique are likely to be restricted
to moderate zenith angles, depending on wavelength and seeing.

\subsubsection{Astrophysical Noise}

While the abovementioned error sources can be to some extent dealt
with and reduced, in order to progress toward the goal of a few
$\mu$as precision, astrophysical noise sources (due to the
environment or intrinsic to the target) cannot be easily
minimized, if present.

For example, I show in figure~\ref{gravpull} the comparison
between the gravitational perturbations (as seen along one of the
two axes in the plane of the sky as a function of time)
induced by a 5 $M_J$ and a 0.1 $M_\odot$ companion to
a 1 $M_\odot$ star with periods of 5 yr and 100 yr, respectively, with
the system at a distance $D = 100$ pc from the Sun.
As one can see from the right panel of the Figure, on a
time-scale short compared to the orbital period of the
stellar companion, the astrometric signature of the
planet is superposed to a large effect. The additional signal
in the data could
be easily misinterpreted as an extra proper motion
component or as a significant acceleration,
depending on the orbital characteristics of the long-period
companion and epochs of observation.

As I mentioned earlier, if the orbital characteristics of the
perturbing stellar companion around a target and/or reference
star(s) are known to exist in advance, they can be modelled within
the context of the observable model. The dynamical effect of a
previously unknown stellar companion constitutes otherwise a
significant source of noise that might hamper the reliability of
orbit reconstruction for a newly detected planet.

Such a problem will affect primarily future high-precision
space-borne global astrometric missions such as $Gaia$, which will
not have the luxury to pre-select the list of targets. In the case
of relative astrometry, if feasible, one could require that the
objects composing the local frame of reference not be orbited by
stellar companions inducing unmodelled signatures larger than a
few $\mu$as. A typical strategy to achieve this, adopted for
example for the selection of the grid stars for $SIM$, is to look
for reference objects that are K giants, and pre-select them on
the basis of medium-precision radial-velocity monitoring. In this
case, the typically large distance of these stars (1 kpc) implies
(provided they are not too faint, otherwise photon noise becomes
an issue) a significant suppression of any astrometric signature
that might significantly pollute the potential planetary signal
from the target (e.g., \citealp{gould01}, and references therein).

Another source of astrophysical noise due to the
environment is the presence of a circumstellar disk. The
motion of the center of mass of the disk, provoked by the
excitation of spiral density waves by an embedded planet, induces an
additional wobble in the stellar position, while time-variable,
asymmetric starlight scattering by the disk can introduce
shifts in the photocenter position.

Takeuchi et al. (\citeyear{takeuchi05}) have recently studied these
effects assuming Jupiter-mass planets embedded in gravitationally
stable circumstellar disks around young solar-type stars
at the distance of the Taurus-Auriga star-forming region
($D\simeq 140$ pc). They conclude that the additional
stellar motion caused dynamically by the disk's gravity
is negligible (sub-$\mu$as) with respect to the signature
from the planet ($\sim 36$ $\mu$as if the planet's semi-major
axis is 5 AU). Variable disk illumination can induce peak-to-peak
photocenter variations of order 10-100 $\mu$as, but they
claim that $SIM$ would not be sensitive to such excursions.
Finally, Boss (\citeyear{boss98}) and Rice et al. (\citeyear{rice03c})
have quantified the magnitude of the astrometric displacement induced dynamically
by a marginally unstable disk. They found that in this case
the effect can be as large as $50-100$ $\mu$as, but the typical
time-scale of this perturbation would be of order of
decades, as compared to a few years of observations with
$SIM$ or $Gaia$, thus such source of astrometric noise should not
constitute a major cause of concern.

The last important class of astrophysical noise sources that can cause
shifts in the observed photocenter is not due to stellar environment
but rather intrinsic to the target. Such noise sources include a variety
of surface temperature inhomogeneities such as spots and flares, and
just like disks, they are primarily characteristic of rapidly rotating,
young stellar objects (e.g., \citealp{bouvier95}; \citealp{schuessler96}, and
references therein).

In the context of a study of the effects of the variety
of astrophysical sources of astrometric noise on the
planet detection capabilities of $Gaia$ (Sozzetti et al., in preparation),
I have implemented a numerical model to
calculate the photometric and astrometric
effects of a distribution of spots over the
surface of a rotating star. The model is based
on the analytical theories developed by Dorren (\citeyear{dorren87}),
Eker (\citeyear{eker94}), and Eaton et al. (\citeyear{eaton96}).
It incorporates a broad range of spot
and star parameters, including stellar limb-darkening,
and it allows for the presence of multiple spots of any shape,
including umbra-penumbra effects.

The key result of the numerical analysis (Figure~\ref{spots})
is that a photometric variation in the visual of $\Delta F/F=10\%$
(rms) corresponds to an astrometric variation
of $\sim 3$ $\mu$as (rms) in the position of a 1 $R_\odot$
pre-main sequence star at the distance of Taurus ($D = 140$ pc). 
The magnitude of the spot-induced photocenter motion
on a T Tauri star is thus comparable to the gravitational
effect of a Jupiter-mass object orbiting the star at 0.5 AU
($\sim 5$ $\mu$as). The effect scales with distance
just like the astrometric signature of a planet, thus
for a nearby, less active sun-like star at 10 pc
its magnitude could still be of the same order
(e.g., \citealp{woolf98}), while the amplitude of
the planet-induced stellar motion would be at least
one order of magnitude larger.

However, astrometric signatures induced for example by a few
Earth-mass planet on a 1 AU orbit around a solar-type star
at a distance of 10 pc covered by spots causing a change in
photospheric flux of $\sim 1\%$ could be comparable in size.
The non-uniformity of illumination of the stellar disk
might then jeopardize successful Earth-sized planet detection with
astrometry around the nearest stars, as well the detection of
Jupiter-sized objects in nearby star-forming regions.
Fortunately, large spotted areas on solar-type stars are
relatively uncommon (the Sun itself, at its peak of activity,
is covered by spots for up to $\sim 0.1$\% of its visible surface.
See for example \citealp{allen00}). Furthermore,
the spot-induced photocenter variation has a
period that is strongly correlated to the photometric excursion
as well as the stellar rotation period (of order
of a few days for T Tauri stars, up to several weeks for
solar-type objects). With the help of careful photometric monitoring,
the two sources of astrometric signal might then be
successfully disentangled.

Ultimately, in order to keep environmental and intrinsic
astrophysical noise sources at the few $\mu$as level,
an important cause of concern primarily for giant planet
searches in star-forming regions, it would be beneficial
to avoid stars with large photometric variations
and objects with particularly large, flared disks.

\subsection{Estimation Process}

The estimation process applies the observable
model and noise model to the data.
The estimation process includes several
functions, such as search techniques, hypothesis
testing, and parameter estimation. 
The observable model provides the estimation
process with a parametric description of the expected
data. The estimation process finds the observable
model parameters that best match the data, with
deviations weighted by the noise model. 
The estimation process may be a generalized least-squares
method that takes advantage of the full noise
covariance matrix constructed from the noise model, as I 
briefly describe below.

Suppose we have performed $n$ measurements of the quantity
$y$ collected in the vector {\bf Y}$(y_1,y_2,\dots,y_n)$,
with associated measurement uncertainties
$\mathbf{\Sigma}(\sigma_1,\sigma_2,\dots,\sigma_n)$.
Furthermore, call {\bf X}$(x_1,x_2,\dots,x_p)$
the vector of $p$ unknown quantities that we want to determine.
Let {\bf F}({\bf X}) be the actual functional form of the observable
model. The method of least squares will attempt to find a
solution to the equation
$\mathbf{Y}+\mathbf{\Sigma} = \mathbf{F(x)}$ in terms of the
unknowns in the model.

Under the assumptions that the unknowns are normally distributed
and are sufficiently small, the set of equations can be expanded
to first order in the unknowns. The resulting system of
{\it equations of condition} can be written as:

\begin{equation}\label{eqcond}
\mathbf{\Delta Y} = \mathbf{Y}- \mathbf{F(X)} =
                    \mathsf{D}\mathbf{\delta}+\mathbf{\Sigma},
\end{equation}
with $\mathbf{\delta}(\delta x_1,\delta x_2,\dots,\delta x_n)$
the vector of new unknowns, and $\mathsf{D}$ the design
matrix containing all partial derivatives of the observable
model with respect to the unknowns. In order for the method
to be applicable, the number of equations of condition must be
larger than the number of unknowns, usually {\it at least}
$n > 2p$.

The objective of the least squares technique is to determine
the vector $\mathbf{\delta}$ that minimizes the sum of the
squares of the components of the vector of uncertainties
$\mathbf{\Sigma}$:
\begin{equation}
\sum_{i=1}^n\sigma_i^2 = \mathbf{\Sigma}^\mathrm{T}\mathbf{\Sigma} =
(\mathbf{\Delta Y}-\mathsf{D}\mathbf{\delta})^\mathrm{T}
(\mathbf{\Delta Y}-\mathsf{D}\mathbf{\delta}),
\end{equation}
where the superscript T indicates transposed. The solution is
given by:
\begin{equation}
\mathbf{\delta_0} = (\mathsf{D}^\mathrm{T}\mathsf{D})^{-1}
\mathsf{D}^\mathrm{T}\mathbf{\Delta Y}
\end{equation}

In a weighted least squares solution, Eq.~\ref{eqcond} is
multiplied on both sides by a square $n\times n$ matrix $\mathbf{G}$
containing zeroes except for the elements on the main diagonal,
which are $g_i=\sigma_i^{-1}$. The formal solution now
becomes:
\begin{equation}
\mathbf{\delta_0} = (\mathsf{D}^\mathrm{T}\mathsf{W}\mathsf{D})^{-1}
\mathsf{D}^\mathrm{T}\mathsf{W}\mathbf{\Delta Y} =
\mathsf{C}\mathsf{D}^\mathrm{T}\mathsf{W}\mathbf{\Delta Y},
\end{equation}
with the weight matrix $\mathsf{W} = \mathsf{G}^\mathrm{T}\mathsf{G}$,
and the covariance matrix of the solution
$\mathsf{C} = (\mathsf{D}^\mathrm{T}\mathsf{W}\mathsf{D})^{-1}$.
The formal standard deviations of any unknown $\delta x_i$ in the solution
will be given by the square-root of the diagonal terms of the covariance
matrix $c_{i}$.

The weight matrix maybe generalized in the case where correlations
between the unknowns are present. In this case, non diagonal
terms will not be zero.
This method is probably superior to a more conventional
weighted-least-squares technique, in which difference
observations are formed to cancel correlated errors,
and each difference is assigned a weight commensurate
with the expected uncorrelated portion of the error.

The latter approach fails to take full
advantage of the knowledge of the correlations, and
residual correlated errors between successive differences
persist. With a more rigorous approach, not
only one can take into account any correlations that
might exist, regardless of their temporal or angular
scale and dependence on other observing parameters,
but the correlation itself becomes part of the solution,
and the self-consistency of the solution can be
determined to establish to what extent the noise
model is satisfactory.

If the functional form $\mathbf{F}(\mathbf{X})$ is
non-linear, as is often the case, multiple iterations of
the linearized least squares procedure must be carried out.
In this case, the quality of the solution ultimately obtained, i.e.
how close the minimum of the functional form adopted for the observable
model is to the real one, will strongly depend on the
point at which the equations are linearized.
Good starting guesses of the parameters of the model would be highly
desirable in order to favor the convergence of the iterative procedure.
To this end, several techniques could be adopted,
such as local and global minimization strategies, including
the simplex method, simulated annealing, or genetic algorithms
(e.g., \citealp{press92}).

Once a solution for the vector of parameters $\mathbf{\delta_0}$
is obtained, it is necessary to assess whether the observable
model employed is indeed representative of the reality.
To this end, a number of tests can be conducted.

The general procedure consists in defining a test
statistic ($\chi^2$ or its transformation $F2$, Fisher's $F$,
Kolmogorov-Smirnov's $D$) which is some function of the
data measuring the distance between the hypothesis and
the data, and then calculating
the probability of obtaining data which have a still
larger value of this test statistic than the value observed,
assuming the hypothesis is true. This probability is called
the confidence level. Small probabilities
(say, less than one percent) indicate a poor fit.
Especially high probabilities (close to one) correspond to
a fit which is too good to happen very often, and may indicate
a mistake in the way the test was applied, such as treating
data as independent when they are correlated.

For the purpose of planet detection, upon rejection of an observable
model which assumes the star is single by means of a goodness-of-fit
test, observations residuals should be inspected for the
presence of hidden periodicities in the measurements.
A possible approach is as follows.

First, a period search is conducted. To this end, the Lomb-Scargle
periodogram analysis could be performed (e.g.,
\citealp{lomb76}; \citealp{scargle82}; \citealp{horne86}).
Alternatively, the observable model could be expanded
to incorporate a perfectly sinusoidal term (circular
orbit), and a star+circular orbit fit performed adopting a dense
grid of trial periods. It should then be possible to test if the new observable model
(star+circular orbit) provides a significant improvement with respect to
the single-star model. For example, a statistical test of
the goodness-of-fit of the single-star and star+circular orbit
model could be adopted, such as the likelihood-ratio test.

If the star+circular orbit fit performs significantly better in a
measurable way, the observable model is then further expanded, to
include a full Keplerian orbit. In the presence of multiple planetary
signals, the procedure is carried out until not further periodicities
can be uncovered and the observation residuals are fully consistent with
noise.

\subsection{Ground-Based Experiments}

The confirmation of theoretical predictions that astrometry
with ground-based monopupil telescopes is limited to the mas precision regime
has been given by numerous experiments. Gatewood (\citeyear{gatewood87}) derived
$\sigma_\mathrm{atm}\approx 3$ mas/h with a reference frame of
$10^\prime-20^\prime$. Han (\citeyear{han89}) showed that a 1 mas/h precision
can be reached for a double star with $\vartheta = 1^\prime$.
Similar conclusions were derived by Gatewood (\citeyear{gatewood91})
and Dekany et al. (\citeyear{dekany94}).

If instead of a single reference star, a dense star field is
used as a frame of reference, the situation improves somewhat.
The theoretical predictions of
$\sigma_\mathrm{atm}\propto A^{-1}\vartheta^{4/3} t^{-1/2}$
(\citealp{lindegren80}) and
$\sigma_\mathrm{atm}\propto A^{-3/2}\vartheta^{11/6} t^{-1/2}$
(\citealp{lazorenko02}) have been substantially confirmed by Pravdo \&
Shaklan (\citeyear{pravdo96}), who demonstrated $\sigma_\mathrm{atm}\approx 150$ $\mu$as/h
with the 5-m Palomar telescope using 15 reference objects in a field
of 90 arcsec. These authors showed also that $\sigma_\mathrm{DCR}\approx 60-100$
$\mu$as within 1 hour of meridian crossing
and at declinations within $45^\circ$ of the zenith.

It is worth noting that, assuming apodization of the entrance pupil and
enhanced symmetrization of the reference field, achieved assigning
a specific weight to each reference star, Lazorenko \& Lazorenko
(\citeyear{lazorenko04})
have recently generalized the theoretical expression for the
astrometric error due to the atmosphere:
\begin{equation}
\sigma_\mathrm{atm} \approx A^{-k/2+1/3}\vartheta^{k\mu/2}t^{-1/2},
\end{equation}

with $2\leq k\leq \sqrt{8N_r+1}-1$, limited by the number $N_r$ of reference
objects, and $\mu\leq 1$ a term dependent on $k$ and the magnitude
and distribution on the sky of the field stars. The classic result
by Lindegren (1980) is recovered for $N_r = 1$. However,
the Lazorenko \& Lazorenko (\citeyear{lazorenko04}) expression predicts
$\sigma_\mathrm{atm}\approx 10-60$ $\mu$as (depending on stellar
field density) for a 10-m telescope, very good seeing (FWHM = 0.4 arcsec),
and $t = 600$ sec. This is about a factor 2-5 improvement with respect
to the prediction of Eq.~\ref{linde}, which would have the goal
of $\sigma_\mathrm{atm}\approx 60$ $\mu$as reached in $t =1 $ hr, for
$A = 10$ meters. The improvement due to this new approach
to the astrometric measurement process (which however neglects
DCR effects) still awaits experimental confirmation.

The promise of long-baseline optical/infrared interferometry for
high-precision astrometry has been tested by a number of experiments
in the past. The Mark-III and NPOI interferometers have achieved
long-term wide-angle astrometric precision at the 10 mas level
(\citealp{hummel94}).
Short-term astrometric performance at the 100 $\mu$as level has been demonstrated
with Mark-III and PTI (\citealp{colavita94}; \citealp{shao99};
\citealp{lane00}), for
moderately close (30 arcsec) pairs of bright ($m_v\approx 2-5$) stars. 
Recently, Lane \& Muterspaugh (\citeyear{lane04}) have demonstrated that 10-$\mu$as
short-term very narrow-angle astrometry is possible for a sample of close,
sub-arcsec binaries observed with PTI in phase-referencing mode.

The predicted astrometric performances of Keck-I (\citealp{boden99})
and VLTI (\citealp{derie03}) will presumably reach the actual limits of
this technique from the ground (unless such an instrument is built at
the South Pole). The two instruments have quoted limiting magnitudes
in the near-infrared (2-2.4 $\mu$m) of $m_k\sim 17-18$ for narrow-angle
astrometry at the 30 $\mu$as and 10-20 $\mu$as level, respectively,
between pairs of objects separated by $< 20-30$ arcsec.
\footnote{For astrometric
planet searches conducted with these instruments, the probability of finding
a reference star with $m_k = 13$, or fainter within 20-30 arcsec from a
target object is about 50\%-60\%, or greater, if the target is not very
far from the galactic plane (\citealp{derie03}).}

\subsection{Space-Borne Experiments}

Relative, narrow-angle astrometry from space has been performed so
far with the Fine Guidance Sensors aboard $HST$, while global
astrometric measurements have been carried out for the
first time by $Hipparcos$.

For $HST$/FGS, the data reduction of the two-dimensional
interferometric measurements entails a number of {\it ad hoc} calibration
and data reduction procedures to remove a variety of random and systematic error
sources from the astrometric reference frame (e.g., \citealp{taff90};
\citealp{bradley91}; \citeauthor{benedict94} \citeyear{benedict94},
\citeyear{benedict99}, and references therein). 
The calibration of random and systematic, long- and short-term
error sources for $HST$/FGS includes removing intra-observation
spacecraft jitter, compensating for temperature variations and
temperature-induced changes in the secondary mirror position,
applying constant and time-dependent optical field angle distortion 
calibrations, correcting for intra-orbit drift, and applying lateral
color corrections during the orbit-to-orbit astrometric modeling
(e.g., \citeauthor{benedict94} \citeyear{benedict94}, \citeyear{benedict99}).

The global single-measurement
error budget for $HST$/FGS astrometry with respect to
a set of reference objects near the target (within the $5\times 5$
arcsec instantaneous field of view of FGS) had received a
pre-launch estimate of $\sim 2.7$ mas by Bahcall \& O'Dell (\citeyear{bahcall80}).
Benedict et al. (\citeyear{benedict94}, \citeyear{benedict99}) confirmed the overall
performance level of the instrument, with single-measurement
uncertainties of 1-2 mas down to $m_v = 16$. The limiting factor
is the spacecraft jitter. A single-measurement
precision below 0.5-1 mas is out of reach for $HST$/FGS.

For $Hipparcos$, a calibration and iterative reduction procedure in five main
steps (\citeauthor{lindekov89} \citeyear{lindekov89}) is 
devised in order to derive values
of positions, proper motions, and parallaxes simultaneously
for $\sim 120,000$ stars by combining one-dimensional angular
measurements along the satellite's instantaneous scanning direction
into a global astrometric solution over the whole celestial sphere. 
These steps include: 1) the determination of the satellite attitude;
2) the estimation of stellar coordinates relative to the main
focal grid; 3) the reference great circle (RGC) reduction, to
determine the abscissae of stars on each RGC; 4) the sphere solution,
to determine the correction to the great circle origins using a
set of instrumental calibration parameters (including chromatic
terms); 5) the determination of the five astrometric parameters
with respect to the rigid reference sphere of RGCs, using all
RGC abscissae, the RGC origins, and instrumental parameters
(\citeauthor{lindekov89} \citeyear{lindekov89}).

Typical uncertainties on the RGC abscissae are of order
of 1.0 mas for bright objects ($m_v < 7$), and degrade up to
$\sim 4.5$ mas for $m_v\geq 11$ (e.g., \citealp{kovalevsky02}).
\footnote{The errors on the final astrometric parameters are not only a
function of $m_v$, but also of the ecliptic latitude $\beta$, a normal
consequence of the adopted scanning law (stars at low latitudes
were observed significantly less often).} These agree well with
pre-launch predictions by Lindegren (\citeyear{lindegren89}).
Without the presence of the atmosphere, and
similarly to $HST$/FGS, the best-achievable single-measurement
precision is limited by the uncertainties in the determination
of the along-scan attitude.

The ability to suppress systematics by at least two orders of
magnitude for a space-borne instrument is a major technological
goal. Both $SIM$ and $Gaia$ promise to achieve this level of
astrometric precision. For the purpose of planet detection with
$SIM$, in order to deliver 1 $\mu$as narrow-angle astrometry in 1
hr integration time down to $m_v \simeq 11-12$, an accuracy on the
position of the delay line of 50 pm with a 10 m baseline must be
achieved (\citealp{shaklan98}). Furthermore, a positional
stability of internal optical pathlengths of $\sim 10$ nm is
required, in order to ensure maintenance of the fringe visibility
(\citealp{neat98}). For a $Gaia$-like instrument, the success in
meeting the goal of $\approx 5-10$ $\mu$as single-measurement
astrometric precision to hunt for planets around bright stars
($m_v < 11-12$) will depend on a) the ability to attain CCD
centroiding errors not greater than 1/1000 of a pixel in the
along-scan direction (\citealp{gai01}) and b) the capability to
limit instrumental uncertainties (thermo-mechanical stability of
telescope and focal plane assembly, metrology errors in the
monitoring of the basic angle) and calibration errors
(chromaticity, satellite attitude, focal plane-to-field
coordinates transformation) down to the few $\mu$as level (e.g.,
\citealp{perryman01}).

\section{Planet Detection with Astrometry: Past and Present Efforts}

Astronomers have long sought to find astrometric perturbations in a
star's motion due to orbiting planet-sized companions. Many attempts
have failed, some have produced more or less significant upper limits,
a few have been successful. I review in turn the history of these efforts.

\subsection{Unfinished Tales: Barnard's Star and Lalande 21185}

During the 1960s, based on the analysis of over 2000 photographic plates
of the Sproul Observatory covering 24 years (1938-1962)
van de Kamp (\citeyear{vandekamp63}, \citeyear{vandekamp69a},
\citeyear{vandekamp69b}) announced the discovery of
perturbations in the motion of Barnard's Star (GJ 699) that could be
explained initially with the presence of a 1.6-1.7 $M_J$ planet on a 24-25
years, eccentric orbit, and then instead in terms of two Jupiter-sized
objects on coplanar, circular orbits with periods of 11.5 and 22 years,
respectively. Through the years, van de Kamp refined his results, extending
the time duration of the photographic plate observations up to 43 years
(1938-1981), and publishing two more papers (\citealp{vandekamp75},
\citeyear{vandekamp82}).
In his last interview on the subject (\citealp{schilling85}), he still claimed that
Barnard's Star was orbited by two massive planets of 0.7 $M_J$ and 0.5 $M_J$,
co-revolving on circular, coplanar orbits with periods of 12 and 20 years,
respectively.

Neither planet was ever confirmed, however. Initial claims by
Jensen \& Ulrych (\citeyear{jensen73}), that observations were compatible with the
presence of up to five planets, were not verified by Gatewood \&
Eichhorn (\citeyear{gatewood73}), who could not detect any additional motion perturbing
Barnard's Star. The existence of giant planets orbiting the star
was further cast in doubt by Hershey (\citeyear{hershey73}), and Heintz
(\citeyear{heintz76}), who explained van de Kamp's results in terms of a number
of unrecognized systematics, including telescope internal motions due to two
phases of cleaning and remounting of the telescope lens 25 years after he
began his observations. Years later Frederick \& Ianna (\citeyear{frederick80}),
Harrington (\citeyear{harrington86}),
and Harrington \& Harrington (\citeyear{harrington87})
reported other independent studies of Barnard's star,
in which no wobble was detected, although van de Kamp's results were
not totally discounted. Croswell (\citeyear{croswell88}), on the other hand, addressed
again the issue of the misinterpretation of incorrect Sproul Observatory data,
concluding that unknown telescope systematics were the more likely
explanation. 

In a more recent study, Gatewood (\citeyear{gatewood95}) ruled out the
presence of massive planets or brown dwarfs ($M_p > 10$ $M_J$) around
Barnard's star, while no conclusion was reached on the existence of
objects of order of the mass of Jupiter or smaller. Using $HST$/FGS astrometry,
Benedict et al. (\citeyear{benedict99}) ruled out the presence of Jupiter-mass
planets with orbital periods $P < 3$ years, but their observations were not
taken for a sufficiently long amount of time to address the period range of
the putative planets discovered by van de Kamp. Schroeder et al.
(\citeyear{schroeder00}) conducted a photometric study of Barnard's Star, and did not
find any supporting evidence for the presence of massive planets and
brown dwarfs at large orbital radii, in agreement with Gatewood's 
(\citeyear{gatewood95}) findings. The latest study of this star has been 
undertaken by K\"urster et al. (\citeyear{kurster03}) using precision 
radial-velocity measurements, which 
allowed them to rule out the presence of planets down to the terrestrial 
mass regime (a few $M_\oplus$) for objects within 1 AU. 
Although studies of Barnard's Star have spanned over half a century, 
no definitive confirmation or disproval has been established.

The possible existence of a giant planet companion (at least several
Jupiter masses) to Lalande 21185 (HD 95735) was first discussed by
Lippincott (\citealp{lippincott60a}, \citeyear{lippincott60b}) on the basis of
photographic plates covering a time-span of 47 years taken with the Sproul telescope.
Gatewood (\citeyear{gatewood74})
did not find any evidence of a planetary signature at the suggested 8-yr
period, but Hershey \& Lippincott (\citeyear{hershey82}) claimed the planetary mass
companion did exist, although on a different, longer period. Based on a limited
dataset covering four years, Gatewood et al. (\citeyear{gatewood92}) were not
able to detect any significant perturbation in the star's proper motion. 
However, four years later Gatewood (\citeyear{gatewood96}), in examining 50 years
or radial-velocity data of Lalande 21185, as well as a more sophisticated
set of astrometric observations, first concluded that the star is indeed
orbited by a 2.0 $M_J$ planet at $\sim 10$ AU, and then suggested the
existence of two giant planets, the second body being less massive than
Jupiter and orbiting at around 3 AU from the parent star. Also in this
case, an independent confirmation has yet to be made on either
of the two planets.

\subsection{Upper Limits and Controversial Mass Determinations}

Prompted by the success of Doppler surveys for giant planets of
nearby stars and by the need to find a method to break the $M_p-i$ degeneracy
intrinsic to radial-velocity measurements, several authors have
re-analyzed in recent years the $Hipparcos$
Intermediate Astrometric Data (IAD), in order to either detect
the planet-induced stellar astrometric motion of the bright hosts, most of which
had been observed by the satellite, or place upper limits to the
magnitude of the perturbation, in the case of no detections. The $Hipparcos$
IAD have been re-processed alone, or in combination with either the
spectroscopic information or with additional ground-based
astrometric measurements.

The first such analysis was performed by
Perryman et al. (\citeyear{perryman96}), who failed to detect the astrometric motion of the
first three planet-bearing stars announced, 51 Peg, 47 Uma, and 70 Vir.
Based on the size of the astrometric residuals to a single-star fit,
they report upper limits on companions masses in the sub-stellar
regime (7-65 $M_J$, depending on confidence levels) for 47 UMa and 70 Vir, while
limits on the companion to 51 Peg, due to its very short period, are
less stringent.

Orbital fits to the $Hipparcos$ IAD can be performed by using the values
of $P$, $\tau$, $e$, and $\omega$ derived from spectroscopy and by solving
for $a$, $i$, and $\Omega$, with the additional constraint
(\citealp{pourbaix00}):

\begin{equation}\label{constraint}
\frac{a\sin i}{\pi_\star} =
\frac{P K\sqrt{1 - e^2}}{2\pi\times 4.7405}
\end{equation}

With this approach, Mazeh et al. (\citeyear{mazeh99}) and
Zucker \& Mazeh (\citeyear{zucker00}) published
astrometric orbits for the outermost planet in the $\upsilon$ And system
and for the planet orbiting HD 10697. The derived semi-major axes of
$1.4\pm 0.6$ mas and $2.1\pm 0.7$ mas, respectively, imply companion
masses of $10.1^{+4.7}_{-4.6}$ $M_J$ and $38\pm 13$ $M_J$, respectively.
These values depart significantly from the minimum masses from
spectroscopy, as a consequence of the small inclination angles
obtained by the fitting procedures ($i = 24^\circ$ and 
$i = 10^\circ$, respectively).

However, two subsequent studies by Gatewood et al. (\citeyear{gatewood01})
and Han et al. (\citeyear{han01}) contributed to spark a controversy over the
reliability of the determination of sub-stellar companion
masses with milli-arcsecond astrometry. In the first work,
Gatewood et al. (\citeyear{gatewood01}) combined the $Hipparcos$ IAD with Multichannel
Astrometric Photometer (MAP; \citeauthor{gatewood87} \citeyear{gatewood87})
observations of $\varrho$ CrB in an astrometric orbital solution that yielded a semi-major
axis of $1.66\pm 0.35$ mas, an inclination of $0^\circ.5$, and a derived
companion mass of $0.14\pm 0.05$ $M_\odot$. The follow-up paper by
Han et al. (\citeyear{han01}) presented $Hipparcos$-based preliminary astrometric
masses for 30 stars with at least one spectroscopically detected
giant planet. The main conclusion of this work is that a significant
fraction ($\sim 40\%$) of the planet candidates are instead stars,
and the remainder sub-stellar companions are in most cases brown dwarfs rather
than planets. The results stem from the derivation of a vast majority
of quasi-face-on orbits, with 60\% of the sample having $i < 5^\circ$
and $27\%$ having $i < 1^\circ$.

On the one hand, if orbits are isotropically oriented
in space, the probability of finding one with $i < 1^\circ$ is
$\approx 1\times 10^{-4}$, thus Han et al. (\citeyear{han01}) come to the conclusion
that the sample of planet-bearing stars is severely biased towards
small inclination angles. On the other hand, rather than having to reject
the planet hypothesis for a substantial fraction of the Doppler candidates,
the systematically very small inclination angles,
and thus very large actual companion masses, could arise as
an artifact of the fitting procedure.
This thesis was indeed put forth by Pourbaix (\citeyear{pourbaix01}) 
and Pourbaix \& Arenou (\citeyear{pourbare01}) (and later 
by Zucker \& Mazeh (\citeyear{zucker01b})). Using different statistical approaches aimed
at assessing the robustness of the derived $Hipparcos$ astrometric orbits,
these authors
demonstrated that the $Hipparcos$ IAD do not have enough precision to actually
reject the planet hypothesis in essentially all cases (although a few border line
cases do exist). 
Thus, essentially all the preliminary astrometric masses derived for stars with
planets observed with $Hipparcos$ (\citeauthor{mazeh99} \citeyear{mazeh99};
\citeauthor{zucker00} \citeyear{zucker00}; \citeauthor{gatewood01} \citeyear{gatewood01};
\citeauthor{han01} \citeyear{han01}) do not survive close statistical scrutiny.

The $Hipparcos$ IAD can still be used however to put
upper limits on the size of the astrometric perturbations, as done by
Perryman et al. (\citeyear{perryman96}) and by Zucker \& Mazeh
(\citeyear{zucker01b}), who could rule out
at the $\sim 2-\sigma$ level the hypothesis of low-mass stellar companions
disguised as planets for over two dozen objects. This, combined with the
fact that the same analysis of $Hipparcos$ data reveals that instead a
significant fraction of the proposed brown dwarf companions from spectroscopy
is stellar in nature, is interpreted as further evidence of the existence
of the brown dwarf desert that separates stellar and planetary mass secondaries.

In the end, the only {\it firm} upper limits on the mass of a
spectroscopically detected extrasolar planet are those placed by McGrath et al.
(\citeyear{mcgrath02}, \citeyear{mcgrath04}) who failed to reveal astrometric motion
of the $M_p\sin i = 0.88$ $M_J$ object on a 14.65-day orbit in the
$\varrho^1$ Cnc multiple-planet system using $HST$/FGS astrometry.
With a nominal single-measurement precision of 0.5 mas, the
failed attempt at detecting any reflex motion in the data implies
that the 1.15 mas preliminary $Hipparcos$-based mass estimate by
Han et al. (\citeyear{han01}) is ruled out at the 3-5 $\sigma$ level, thus establishing
an updated mass upper limit of $\sim 30$ $M_J$ and firmly confirming that
the object is sub-stellar in nature.

\subsection{Actual Measurements and Work in Progress}

It was not until two years after the first confirmation of the planetary
nature of the companion to HD 209458 via detection of its transits 
across the disk of the parent star (\citeauthor{charbon00} 
\citeyear{charbon00}; \citealp{henry00})
that astrometric techniques finally provided the first undisputed value of
the actual mass of a Doppler-detected planet. Narrow-field relative astrometry
of the multiple-planet host star GJ 876 was carried out by Benedict et al.
(\citeyear{benedict02}) using $HST$/FGS.

The goal of this project was to determine the astrometric wobble induced
on the parent star by the outer planet. At a distance $D = 4.7$ pc, and with a
nominal primary mass of the M4 dwarf star $M_\star = 0.32$ $M_\odot$, the
planet with a projected mass $M_p\sin i\sim 2$ $M_J$ on a $P = 60$ days
orbit was predicted to produce a minimum gravitational perturbation of
$\sim 270$ $\mu$as, which was deemed detectable by the typical 0.5 mas
single-measurement precision of $HST$/FGS. 
Benedict et al. (\citeyear{benedict02}) utilized five reference stars within a 
few arc-minutes 
from the target, and derived the perturbation size, inclination angle, and
mass of GJ 876b from a combined fit to the available astrometry and spectroscopy.
They found $\alpha = 250\pm 60$ $\mu$as, $i=84^\circ\pm 6^\circ$, and
$M_p = 1.89\pm 0.34$ $M_J$.

In the recent announcement (\citealp{mcarthur04}) of the discovery of
a Neptune-sized planet on
a 2.8 days orbit in the $\varrho^1$ Cnc system (which brought the number
of planets in the systems to a total of four), $HST$/FGS astrometry
played again an important role. The authors in fact re-analyzed the
available data on $\varrho^1$ Cnc which had allowed
McGrath et al. (\citeyear{mcgrath02}, \citeyear{mcgrath04})
to put stringent upper limits on the mass of the 14.65-day period planet,
and estimated, from the small arc of the orbit covered in the limited
$HST$ dataset, a perturbation size ($1.94\pm 0.4$ mas) and inclination
($53^\circ\pm6^\circ.8$) for the outermost planet, orbiting at $\sim 5.9$ AU.
Under the assumption of perfect coplanarity of all planets in the
system, this implies an actual mass for the innermost planet of
$17.7\pm 5.57$ $M_\oplus$.

Currently, Benedict et al. (\citeyear{benedict03a}, \citeyear{benedict03b},
\citeyear{benedict04}) are monitoring with
$HST$/FGS the stars $\upsilon$ And and $\varepsilon$ Eri, and plan to
combine the data with the available radial-velocity datasets and with
lower-per-measurement precision ground-based astrometry. The predicted
minimum perturbation sizes of the long-period (3.51 yr and 6.85 yr,
respectively) planets orbiting these stars
($\alpha_{\upsilon\,\mathrm{And}}\simeq 540$ $\mu$as and
$\alpha_{\varepsilon\,\mathrm{Eri}} \simeq 1120$ $\mu$as, respectively)
should be clearly detectable with $HST$/FGS, provided a sufficient
time baseline for the observations.

\section{Future Prospects}

A number of authors have tackled the problem of
evaluating the sensitivity of the astrometric technique required to 
detect extrasolar planets and reliably measure their orbital elements and 
masses. In particular, the works by Casertano et al. (\citeyear{caser96}), Lattanzi
et al. (\citeyear{latt97}, \citeyear{latt00a}, \citeyear{latt00b},
\citeyear{latt02}, \citeyear{latt05}), and Sozzetti et al. (\citeyear{sozzetti00},
\citeyear{sozzetti01}, \citeyear{sozzetti03a})
were specifically tailored to $Gaia$, those of Casertano \&
Sozzetti (\citeyear{caser99}), Sozzetti et al. (\citeyear{sozzetti02},
\citeyear{sozzetti03b}), Ford \& Tremaine (\citeyear{ford03}),
Ford (\citeyear{ford04}), and Marcy et al. (\citeyear{marcy05}) where instead
centered on $SIM$. Black \& Scargle (\citeyear{black82}) and
Eisner \& Kulkarni (\citeyear{eisner01}, \citeyear{eisner02})
studied the general problem of the detectability of periodic signals with the
astrometric technique alone or in combination with spectroscopic
measurements, while Konacki et al. (\citeyear{konacki02}) and
Pourbaix (\citeyear{pourbaix02})
explored to some extent the reliability of orbit reconstruction of future
astrometric missions when all parameters have to be derived
from scratch, in the limit of high and low signal-to-noise ratios.

The abovementioned exploratory works which provided a first
assessment of the planet detection capabilities of $Gaia$ and $SIM$
adopted a qualitatively correct description of the measurements each
mission will carry out. For $Gaia$, the then-current scanning law
was adopted, while for $SIM$ reference stars and realistic
observation overheads were included.
The authors implemented realistic data analysis techniques
based on both the $\chi^2$ test and periodogram search for
estimating detection probabilities as well as non-linear
least squares fits to the data to determine orbital parameters
and planet masses ranging from 1 $M_J$ down to 1 $M_\oplus$.

From the point of view of data simulation, the major
simplifying assumption of these studies is the
idealization of the adopted instrument. Measurement errors assume
simple gaussian distributions, and knowledge of the spacecraft
attitude is assumed perfect, with no additional instrumental
effects, measurement biases, and calibration imperfections.
In terms of data analysis procedures, the most relevant
simplification is the adoption of perturbations of the true
values of all parameters as initial
guesses for the non-linear fits, largely neglecting the
difficult problem of identifying adequate configurations of
starting values from scratch.

\subsection{Planet Detection}

Detection probabilities are determined based on a $\chi^2$ test
of the null hypothesis that there is no planet. Five-parameter,
single-star fits to the simulated data-sets
are carried out, and observation residuals are
inspected. Residuals large compared to the assumed single-measurement
precision will induce a failure of the $\chi^2$ test, at a
given confidence level.

The two parameters upon which detection probabilities
mostly depend are the astrometric signal-to-noise ratio
$\alpha/\sigma_m$ and the period $P$, while eccentricity and 
orientation in the plane of the sky do not significantly affect 
planet detectability. Figure~\ref{detect} shows
iso-probability contours for $SIM$ as a function of $\alpha/\sigma_m$ and
$P$, based on a $\chi^2$ test with a confidence level of 95\%.

For both instruments, assuming a realistic number of data
points throughout the nominal mission lifetimes $T = 5$ years,
$\alpha/\sigma_m\simeq 2$ is sufficient to detect planetary
signatures for $P\leq T$. As orbital sampling gets increasingly
worse for $P > T$, the required signal rises sharply, especially
for high detection probabilities.
The same qualitative behavior of generic detection curves was
recovered by Eisner \& Kulkarni (\citeyear{eisner01}), who also provided
analytical expressions for the behavior of the astrometric
sensitivity to planetary signatures in the two regimes.

\subsection{Orbit Reconstruction and Mass Determination}

Upon detection of its signal,
the goal of determining a planet's orbital characteristics
and mass requires the adoption of observable models
with at least 12 parameters (5 astrometric + 7 describing
the full Keplerian motion). For $SIM$, the model is further
complicated by the simultaneous solution for the
astrometric parameters of the local frame of reference
(5 for each astrometrically clean reference star).
The simultaneous fit to both astrometric and orbital
parameters strongly reduces the covariance between
proper motion and astrometric signature pointed out by
Black \& Scargle (\citeyear{black82}), in particular for $P\leq T$.

A standard metric to understand how well the observable
model performs on the simulated data is the convergence
probability, i.e. the fraction of the final values of each
parameter that falls within a given fractional error of
the true values. I show in Figure~\ref{orbparmass}
the $Gaia$ convergence probability to 10\% fractional uncertainty
for $a$, $P$, $e$, and $i$ as a function of the distance
from the Sun, for a Jupiter-Sun system with $P\ll T$, $P\simeq T$,
and $P\gg T$.

As a general result, $\alpha/\sigma_m\simeq 5$ is required
for orbit reconstruction and mass determination at the
20-30\% accuracy level, while $\alpha/\sigma_m\simeq 10-15$
is necessary for a more stringent 10\% accuracy requirement.

As it can be seen in Figure~\ref{orbparmass},
orbital periods twice as long as the mission duration
induce significant degradation in the quality of the
orbit reconstruction, although different parameters are
affected differently. For example, the correct period of the
signal is more easily identified as it is independent of the
Keplerian nature of the problem (e.g., \citealp{monet79}). Short
periods also cause a degradation of the results, due to the
increasingly smaller amplitude of the perturbation, an
effect that overruns the increasingly larger number
of orbital revolutions sampled during the mission duration.

Orbital eccentricity also plays a very significant role,
when attempting to obtain an orbital solution. The deterioration
of orbit determination is especially prominent for long-period
planets, where the limited orbital sampling couples with
large values of $e$, with the result that orbits are
increasingly more unlikely to be sampled during pericenter
passage, and the correct orbit size and geometry become very
difficult to identify correctly.

On the other hand, the inclination of the orbital plane
does not impact very significantly the ability to accurately
determine the orbital parameters and mass of a planet, unless
$i\longrightarrow 90^\circ$. In quasi edge-on configurations,
in fact, the projected stellar motion is reduced to one dimension,
and a considerable amount of information is lost. However, this
effect is already negligible for configurations departing
from exactly edge-on by a few degrees (\citeauthor{eisner02} \citeyear{eisner02};
\citeauthor{ford04} \citeyear{ford04}).

Finally, unlike $Gaia$,
$SIM$ will have the leisure to choose the number $N_o$ and
timing of the observations as well as the number $N_r$ of
reference objects. Both detection probabilities
and the quality of orbit determination are sensitive to
these parameters with simple parameterizations given by $\sim \sqrt{N_o}$ and
$\sim \sqrt{N_r}$ (\citeauthor{sozzetti02} \citeyear{sozzetti02}).
Ford (\citeyear{ford04}) has studied in detail a wide range of possible observing
schedules, and concluded that both planet detection and
orbit reconstruction are relatively insensitive to
the specific choice of the distribution of observations.

\subsection{Multiple-Planet Systems}

The limiting ability to detect and characterize planetary systems
with $\mu$as astrometry has been estimated by Sozzetti et al.
(\citeyear{sozzetti01}, \citeyear{sozzetti03b}), utilizing as test-cases the
then-current lists of multiple-planet systems discovered by Doppler surveys. In their
works, the authors neglected any complications deriving from
significant perturbations of the planetary orbits due to
strong planet-planet secular or resonant dynamical interactions.

Under the assumption of sufficient data redundancy with respect to
the number of parameters in the observable model fitted to the
observations\footnote{If $N_\mathrm{pl}$ is the number of planets
in the system, then {\it at least} $N_o > 2\times(5+7\times
N_\mathrm{pl})$ for $Gaia$, and $N_o > 2\times(5+7\times
N_\mathrm{pl}+5\times N_r)$ for $SIM$ is required}, the detection
of additional components in a system will be reliably carried out.
Only border-line cases, in which a signal with
$\alpha/\sigma_m\simeq 1$ is not properly modelled and subtracted,
will produce a significant increase in the false detection rates.
For such cases, and in the limit for $P\leq T$, a period search
would add robustness to the detection method, while the least
squares technique combined with Fourier analysis would arguably be
preferred when attempting to detect signals with $P > T$.

The typical accuracy of multiple-planet orbit reconstruction and
mass determination will be degraded by 30-40\% with respect to the
single-planet case, a relatively modest deterioration particularly
for well-sampled, well-spaced orbits with $\alpha/\sigma_m\geq 10$.

The ability of astrometry to determine the full set of orbital parameters
implies that for favorable multiple-planet configurations it should be
possible to derive a meaningful estimate of the relative inclination
angle (e.g., \citealp{kells42}):

\begin{equation}\label{inclrel}
\cos i_\mathrm{rel} = \cos i_\mathrm{in}\cos i_\mathrm{out}+ \sin
i_\mathrm{in}\sin i_\mathrm{out} \cos(\Omega_\mathrm{out}-
\Omega_\mathrm{in}),
\end{equation}
where $i_\mathrm{in}$ and $i_\mathrm{out}$, $\Omega_\mathrm{in}$
and $\Omega_\mathrm{out}$ are the inclinations and lines of nodes
of the inner and outer planet, respectively.

I show in Figure~\ref{coplan} the estimated accuracy with which
$SIM$ could determine the coplanarity (i.e., $i_\mathrm{rel}\simeq 0.0$)
between pairs of planetary
orbits as a function of the common inclination angle, for 11
known multiple-planet systems.

For configurations in which
all components produce $\alpha/\sigma_m\gtrsim 10$, coplanarity
could be established with typical uncertainties of a few degrees,
for periods up to twice the mission duration. In systems where
at least one component has $\alpha/\sigma_m\longrightarrow 1$,
accurate coplanarity measurements are compromised, and mutual inclinations
can only be determined with uncertainties of several tens of degrees.

Finally, if combined radial velocity + astrometric solutions were
to be carried out on single- or multiple-planet systems, the quality
of orbit reconstruction and mass determination would be significantly
improved, especially in the long period regime ($P > T$) and for
edge-on configurations, while well-sampled, well-measured orbits
($P\leq T$, $\alpha/\sigma_m\gg 1$) would be only marginally
improved by radial velocity + astrometric
solutions (\citeauthor{eisner02} \citeyear{eisner02};
\citeauthor{sozzetti03a} \citeyear{sozzetti03b}).

\subsection{The Search for Good Starting Values}

The convergence of non-linear fitting procedures and the
quality of orbital solutions can both be significantly affected
by the choice of the starting guesses. In the absence of any
kind of {\it a priori} information on the actual presence of
planets around a given target, all orbital parameters will
have to be derived from scratch. The results of, e.g.,
Han et al. (\citeyear{han01}) already provided a word of caution on the
reliability of low $S/N$ astrometric orbits, even when
constraints on some of the parameters are available from
spectroscopy. It is thus crucial to investigate new strategies
in the fitting procedure to maximize the robustness of the
solutions obtained.

Pourbaix (\citeyear{pourbaix02}) tackled the problem in the context of a work
on the precision achievable on the orbital parameters of
astrometric binaries from two- and one-dimensional observations,
in the case of low $S/N$. He proposed a two-dimensional
global grid search approach in the ($e$, $\tau$) space coupled to
a guess on $P$ by means of a period search technique (e.g.,
\citealp{horne86}), while fitting a linearized model
in the four Thiele-Innes elements (e.g., \citealp{green85}).

Konacki et al. (\citeyear{konacki02}) applied a `frequency decomposition' method
to simulated $SIM$ observations of $\upsilon$ And. This approach is
based on a Fourier expansion of the Keplerian motion, in which
the coefficients of the successive harmonics are functions of
all orbital elements. The values of the latter obtained from the
linear least squares solution performed with the Fourier expansion are
then utilized as starting guesses of a local minimization of
the non-linear problem. This method avoids the complications
of a global-search approach in several dimensions, which can
be computationally very intensive. However, the authors
did not attempt to validate their approach in cases departing
from the favorable ($P\leq T$, $\alpha/\sigma_m\gg 1$) configuration
studied.

The most detailed study on this subject is the one currently
carried out by the $Gaia$ Planetary Systems Working Group. 
Lattanzi et al. (\citeyear{latt05}) have recently presented preliminary
results of an on-going, large-scale double-blind test campaign
that has been set up in order to provide a realistic
assessment of the $Gaia$ capabilities in detecting extra-solar
planets.

The double-blind test protocol envisions three distinct
groups of participants. The Simulators
define and generate simulated observations of stars with and
without planets with a $Gaia$-like satellite; the Solvers define
detection tests, with levels of statistical significance
of their own choice, and orbital fitting algorithms, using
any local, global, or hybrid solution method that they devise is best;
the Evaluators compare simulations and solutions and draw a
first set of conclusions on the process.

As an illustrative example,
Figure~\ref{doubleblind} shows one of the results of a simulation of
50,000 stars orbited by a single planet having $0.2\leq P\leq 12$ yr,
and producing $2\lesssim \alpha/\sigma_m\lesssim 100$. The present-day
$Gaia$ scanning-law is utilized, with a single-measurement precision
$\sigma_m = 8$ $\mu$as. The plot
shows how the derived periods by one of the Solvers
compare to the true simulated ones. The
most striking result is the ability to derive very accurate estimates
of the period for $P\leq 6$ yr, for the full range of $\alpha/\sigma_m$
and for all possible values of $0\leq e\leq 1$ and
$0^\circ\leq i\leq 90^\circ$. For periods exceeding the mission
duration by over 20\%, it becomes increasingly difficult to identify
the correct value of $P$. In this case, part of the signal can be 
absorbed in the stellar proper motion, with the net result that the 
size and period of the perturbation are systematically underestimated. 

However, the preliminary findings by Lattanzi et al. (\citeyear{latt05}) show
that `mission-ready' detection and orbital fit packages
(including reliable estimates of the covariance matrix of the
solutions) tailored to future high-precision
astrometric observatories,
requiring  no {\it a priori} knowledge of the orbital elements,
can already achieve good performances.

\section{Discussion: Astrometry in Perspective}

The classic way to gauge the effectiveness of different planet search techniques 
is to compare their respective discovery spaces, defined in terms of the
planets of given mass and period each method will be able to detect.
As an illustrative example,
I show in Figure~\ref{discovery} the $M_p$-$P$ diagram with
the plotted present-day sensitivities of transit photometry and
radial-velocity, and with the expected $SIM$ and $Gaia$ detection thresholds
at 10 pc and 150 pc, respectively. For the radial-velocity detection
curve the simulations by Sozzetti et al. (\citeyear{sozzetti05}) were utilized,
while for $SIM$ and $Gaia$ the Sozzetti et al. (\citeyear{sozzetti02},
\citeyear{sozzetti03a}) results were
used. The sensitivity for transit photometry was derived
based on the Gaudi et al. (\citeyear{gaudi05}) analytical dependence of the detectable
planet radius $R_p$ on $P^{1/6}$ (converted to $M_p\propto P^{1/2}$
assuming constant planet density), under the (naive) hypothesis of
uniform sampling.

Simply taking at face value the curves of Figure~\ref{discovery},
however, can lead to important misunderstandings on the
intrinsic relevance of the different techniques to planetary
science. For example, the sensitivity of photometric techniques
to transiting planets with $P\gtrsim 10$ days is strongly suppressed,
and this detection method is useless if the planet does not transit.
However, the information this technique provides for the very close-in
objects discovered is extremely valuable, and it cannot be provided
by radial-velocity and astrometry.

A more effective way to proceed is then to gauge the relative importance of
different planet detection techniques by looking at their discovery
potential not {\it per se}, but rather
in connection to outstanding questions to be
addressed and answered in the science of planetary systems,
such as those I illustrated at the end of Section 2. I summarize
below some of the most important issues for which $\mu$as
astrometry will play a key role.

\subsection{The hunt for Other Earths}

The holy grail in extra-solar planet science is clearly the
direct detection and characterization of Earth-sized, habitable planets,
with atmospheres where bio-markers (e.g., \citealp{lovelock65};
\citealp{ford01}; \citealp{desmarais02}; \citealp{selsis02}; \citealp{seager05})
might be found that could give clues on the possible
presence of life forms. Imaging terrestrial planets
is presently the primarily science goal of the coronagraphic and
interferometric configurations of the Terrestrial Planet Finder
(TPF; \citealp{beichman02}), and of the Darwin Mission (\citealp{fridlund00}).

Space-borne transit photometry carried out with $Corot$
(\citealp{baglin02}) or $Kepler$ (\citeauthor{borucki03} 
\citeyear{borucki03}) has the
potential to be the first technique to make such a detection.
However, astrometry of all nearby stars within 10-20 pc from the Sun at the $\mu$as
level (with $SIM$ and $Gaia$ in space, and possibly with Keck-I and VLTI
from the ground) will be an essential ingredient in order to be able to
provide Darwin/TPF with a) systems containing {\it bona fide}
terrestrial, habitable planets (\citeauthor{ford03} \citeyear{ford03};
\citeauthor{sozzetti02} \citeyear{sozzetti02}; \citeauthor{marcy05} \citeyear{marcy05}),
and b) a comprehensive database of F-G-K-M stars
with and without detected giant planets orbiting out to a few AU from
which to choose additional targets based on the presence or absence of Jupiter
signposts (\citeauthor{sozzetti03a} \citeyear{sozzetti03a}). Such measurements
will uniquely complement ongoing and planned radial-velocity programs
aiming at $\lesssim 1$ m s$^{-1}$ precision (e.g., \citealp{santos04b}),
and exo-zodiacal dust emission observations from the ground with Keck-I, LBTI, and VLTI.

\subsection{Statistical Properties and Correlations}

As discussed in Sections 2.1 and 2.2, 
planet properties (orbital elements and mass distributions, and correlations
amongst them) and frequencies are likely to depend upon the characteristics of
the parent stars (spectral type, age, metallicity,
binarity/multiplicity). It is thus desirable to be able to provide as large 
a database as possible of stars screened for planets.

The size of the stellar sample screened for planets by an all-sky astrometric
survey such as $Gaia$ (\citeauthor{latt00a} \citeyear{latt00a})
could be of order of a few hundred thousand relatively bright ($m_v < 13$)
stars with a wide range of spectral types, metallicities, and ages
out to $\sim 150$ pc. The sample-size
is thus comparable to that of planned space-borne transit surveys,
such as $Corot$ and $Kepler$. The statistical value
of such a sample is better understood when one considers that,
depending on actual giant planet frequencies as a function of
spectral type and orbital distance, at least a few thousand
planets could be detected and measured (\citeauthor{latt02} \citeyear{latt02}).
This number is comparable to the present-day size of the target lists of
ground-based Doppler surveys. Finally, the ranges of orbital parameters
and planet host characteristics
probed by an all-sky astrometric planet survey would crucially complement both
transit observations (which strongly favor short orbital periods and are
subject to stringent requisites on favorable orbital alignment), and
radial-velocity measurements (which can be less effectively carried out for
stars covering a wide range of spectral types, metallicities, and ages and do not allow
to determine either the true planet mass or the full three-dimensional
orbital geometry).

\subsection{Tests of Giant Planet Formation and Migration}

The competing giant-planet formation models
make very different predictions regarding formation time-scales, planet
mass ranges, and planet frequency as a function of host
star characteristics. Furthermore, correlations between orbital elements
and masses, and possibly between the former and some of the host star
characteristics (metallicity) might reflect the outcome of a variety
of migration processes and their possible dependence on environment (see
Sections 2.1 and 2.2).
These predictions could be tested on firm
statistical grounds by extending planet surveys to large samples of PMS
objects and field metal-poor stars.

The full sample of $\sim 1500$ relatively bright ($m_v < 13$), nearby
($D \lesssim 150-200$ pc), field
metal-poor stars presently known could be screened for giant planets on wide orbits
by $Gaia$ or $SIM$, thus complementing the shorter-period
ground-based spectroscopic surveys (\citeauthor{sozzetti05} \citeyear{sozzetti05}),
which are also limited in the sample sizes due to the intrinsic faintness and
weakness of the spectral lines of the targets. These data combined
would allow for improved understanding of the behavior of the
probability of planet formation in the low-metallicity regime, by
direct comparison between large samples of metal-poor and metal-rich
stars, in turn putting stringent constraints on the proposed planet formation models.
Disproving or confirming the existence of the $P$-[Fe/H] correlation
would also help to understand whether metallicity plays a
significant role in the migration scenarios for giant planets.

High-precision astrometric measurements of at
least a few hundred relatively bright ($m_v < 13-14$) PMS stars in a
dozen of nearby ($D < 200$ pc) star-forming regions could be carried out
with $SIM$ and $Gaia$, searching for planets orbiting at 1-5 AU.
The possibility to determine the epoch
of giant planet formation in the protoplanetary disk would
provide the definitive observational test to distinguish
between the proposed theoretical models.
These data would uniquely complement near- and
mid-infrared imaging surveys (e.g., \citealp{burrows05}, and references
therein) for direct detection of young, bright, wide-separation
($a > 30-100$ AU) giant planets.

\subsection{Dynamical interactions in Multiple-Planet Systems}

The different sources of dynamical interactions proposed to explain the highly
eccentric orbits of planetary systems (see Section 2.1) give rise to
significantly different orbital alignments.
An effective way to understand their relative roles would
involve measuring the relative inclination angle between pairs
of planetary orbits. Studies addressing the long-term dynamical stability
issue for multiple-planet systems, as well as the possibility of formation
and survival of terrestrial planets in the Habitable Zone of the
parent star (see Section 2.3), would also greatly benefit from knowledge of whether
pairs of planetary orbits are coplanar or not.

The only way to provide meaningful estimates of the full
three-dimensional geometry of {\it any} planetary system (without
restrictions on the orbital alignment with respect to the line of sight) is
through direct estimates of the mutual inclinations angles using
high-precision astrometry (\citeauthor{sozzetti01} \citeyear{sozzetti01},
\citeyear{sozzetti03b}). For a $Gaia$-like,
all-sky survey instrument, the database of potential targets out
to 50-60 pc is of order of a few tens of thousand objects
(\citeauthor{sozzetti01} \citeyear{sozzetti01}). These data, combined with those
available from Doppler measurements and transit photometry and transit timing
(e.g., \citealp{miralda02}; \citealp{holman05}; \citealp{agol05}),
would then allow to put studies of the dynamical evolution of
planetary systems on firmer grounds.

\subsection{Concluding Remarks}

Despite several decades of attempts, and a few recent successes,
astrometric measurements with milli-arcsecond precision have so far
proved of limited utility when employed as a tool to search for
planetary mass companions orbiting nearby stars.
However, an improvement of 2-3 orders
of magnitude in achievable measurement precision, down to the few $\mu$as
level, would allow this technique to achieve in perspective the same
successes of the Doppler method, for which the improvement from the km s$^{-1}$
to the few m s$^{-1}$ precision opened the doors for ground-breaking
results in planetary science.

In this paper I have reviewed a series of technological,
statistical, and astrophysical issues that future ground-based as
well as space-borne efforts will have to face in their attempts to
discover planets. At the $\mu$as precision level, independently on
the type of instrument utilized (either filled- or diluted
aperture telescopes), a number of important modifications to the
standard definition of astrometric observable (the stellar
position in the instrument-specific reference frame) will have to
be introduced, such as subtle effects due to general relativity.
Astrophysical noise sources will have to be taken into account,
which may mimic the presence of a planet, such as significant
stellar surface activity. Several tools will have to be considered
when attempting to derive reliable orbital solutions, such as
optimized strategies to find good initial configurations for the
orbital parameters. However, the greatest challenge will be to
build instrumentation, both from ground (Keck-I, VLTI) and in
space ($SIM$ and $Gaia$), capable of attaining the technologically 
demanding requirements to achieve a targeted single-measurement
precision $\sigma_m \simeq 1-10$ $\mu$as. Provided these will
be met, astrometry during the next decade bears the potential to
provide critical contributions to planetary science, which are
crucially needed in order to complement the expectations from
other indirect and direct planet detection methods, and refined
theoretical understanding, for continuous improvements in the
field of the formation and evolution of planetary systems.

\acknowledgments

During the preparation of this manuscript,
I have benefited from very fruitful discussion with
many colleagues. I am particularly grateful to A. P. Boss, S. Casertano,
D. Charbonneau, D. W. Latham, M. G. Lattanzi, G. W. Marcy, M. Mayor,
D. Pourbaix, D. Queloz, D. D. Sasselov, and G. Torres for helpful comments 
and insights. I thank an anonymous referee for a very critical revision of 
an earlier version of this paper, and for illuminating suggestions and 
recommendations that greatly improved the manuscript.
The author acknowledges support from the Smithsonian
Astrophysical Observatory through the SAO Predoctoral Fellowship
program and the Keck PI Data Analysis Fund (JPL 1262605).
This research has made use of NASA's Astrophysics Data System Abstract Service
and of the SIMBAD database, operated at CDS, Strasbourg, France.

\clearpage

\figcaption[]{The magnitude of changing parallax and perspective
acceleration as a function of stellar radial and tangential
velocity\label{changepimu}}

\figcaption[]{Left: the size of the gravitational perturbation,
along one axis, induced on a 1 $M_\star$ star at 100 pc
by a 5 $M_J$ planet (solid line) with $P = 5$ yr,
and a 0.1 $M_\odot$ star on a circular (dotted line)
and eccentric ($e = 0.8$, dashed-dotted line) orbit with $P = 100$ yr.
Right: a zoomed-in region covering 5 years of observations
\label{gravpull}}

\figcaption[]{The relationship between the photometric variation
$\Delta F/F$ in the visual and the photocenter shift due to starspots
on a T Tauri star at 140 pc\label{spots}}

\figcaption[]{Iso-probability contours for planet detection with
$SIM$ at the 95\% confidence level. The results are expressed
as a function of $P$ and $\alpha/\sigma_m$, and averaged
over all orbital parameters. The dashed lines indicate
the equivalent signature at the given distance for a system
composed of a solar-mass primary and a 20 $M_\oplus$ planet
(1 $M_J$ in parentheses).
From Sozzetti et al. (\citeyear{sozzetti02}); copyright Astronomical
Society of the Pacific; reproduced with permission\label{detect}}

\figcaption[]{$Gaia$ Convergence probability to a 10\% accuracy for
$a$, $P$, $e$, and $i$, as a function of the distance from
the Sun. A Sun-Jupiter systems is assumed. Symbols of
different shapes correspond to different periods: triangles
for 0.5 yr, diamonds for 5 yr, and crosses for 11.8 yr.
From Lattanzi et al. (\citeyear{latt00a}); copyright Royal Astronomical Society;
reproduced with permission\label{orbparmass}}

\figcaption[]{Mutual inclination $i_\mathrm{rel}$ between
pairs of planetary orbits, as a function of the common
inclination angle with respect to the line of sight,
for 11 multiple-planet systems measured by $SIM$.
In each panel, the corresponding
uncertainties are computed utilizing the formal expressions
from the covariance matrix of the multiple Keplerian fit.
From Sozzetti et al. (\citeyear{sozzetti03b}); copyright Astronomical
Society of the Pacific; reproduced with permission\label{coplan}}

\figcaption[]{True vs. derived orbital periods
(in yr) for a set of 50,000 orbital solutions in the
context of the Double-blind Test Campaign for planet detection
with $Gaia$\label{doubleblind}}

\figcaption[]{Planet discovery space in the $M_p-P$ diagram for
different techniques. Detection curves are defined on the basis
of a 3-$\sigma$ criterion. For $SIM$ and $Gaia$, $\sigma_m = 2$ $\mu$as
$\sigma_m = 8$ $\mu$as are assumed, respectively.
For radial-velocities, $\sigma_m = 3$ m/s. Finally, for transit
photometry, $\sigma_m = 5$ mmag. The time-span of the
observations is set to $T = 5$ yr.
\label{discovery}}

\clearpage

\begin{deluxetable}{@{}cc}
\tablecaption{Comparison between orders of magnitude of parallax, proper
motion, and astrometric signatures induced by planets of various
masses and different orbital radii.
A 1-$M_\odot$ star at 10 pc is assumed\label{alpha}}
\tablewidth{0pt} \tablehead{\colhead{Source} & \colhead{$\alpha$}}
\startdata
Jupiter@1 AU  & 100 $\mu$as \\
Jupiter@5 AU & 500 $\mu$as \\
Jupiter@0.05 AU & 5 $\mu$as  \\
Neptune@1 AU & 6 $\mu$as \\
Earth@ 1 AU & 0.33 $\mu$as  \\
Parallax & $1\times 10^5$ $\mu$as  \\
Proper Motion & $5\times 10^5$ $\mu$as/yr \\
\enddata
\end{deluxetable}

\clearpage

\begin{deluxetable}{@{}ccc}
\tablecaption{Relativistic light deflection effects of various solar-system
objects. The post-Newtonian angular displacement $\alpha$
of a limb-grazing light ray due to the gravitational field of
various solar-system objects assumes the latter are perfect spheres. The angle
$\delta_\mathrm{min}$ is the angular distance between the body and the
light ray from a distant source at which the effect is still equal to 1 $\mu$as.
For Earth and the Moon, two values are given: for a geostationary observer and
for a satellite at a distance of $10^6$ km from Earth\label{deflect}}
\tablewidth{0pt} \tablehead{\colhead{Source} &
\colhead{$\alpha$ ($\mu$as)} & \colhead{$\delta_\mathrm{min}$(1 $\mu$as)}}
\startdata
Sun  & $1.75\times 10^6$ & $180^\circ$ \\
Mercury & 83 & $9^\prime$ \\
Venus & 493 & $4^\circ.5$  \\
Earth & 574 & $178^\circ/123^\circ$ \\
Moon & 26  & $9^\circ/5^\circ$ \\
Mars & 116  & $25^\prime$  \\
Jupiter & 16270 & $90^\circ$ \\
Saturn & 5780 & $17^\circ$\\
Uranus & 2080 & $71^\prime$\\
Neptune & 2533 & $51^\prime$\\
Ganymede & 35 & $32^{\prime\prime}$\\
Titan & 32 & $14^{\prime\prime}$\\
Io & 31 & $19^{\prime\prime}$\\
Callisto & 28 & $23^{\prime\prime}$\\
Europa & 19 & $11^{\prime\prime}$\\
Triton & 10 & $0.7^{\prime\prime}$\\
Pluto & 7 & $0.4^{\prime\prime}$\\
\enddata
\end{deluxetable}

\clearpage

\begin{figure}
\epsscale{0.80}
\plotone{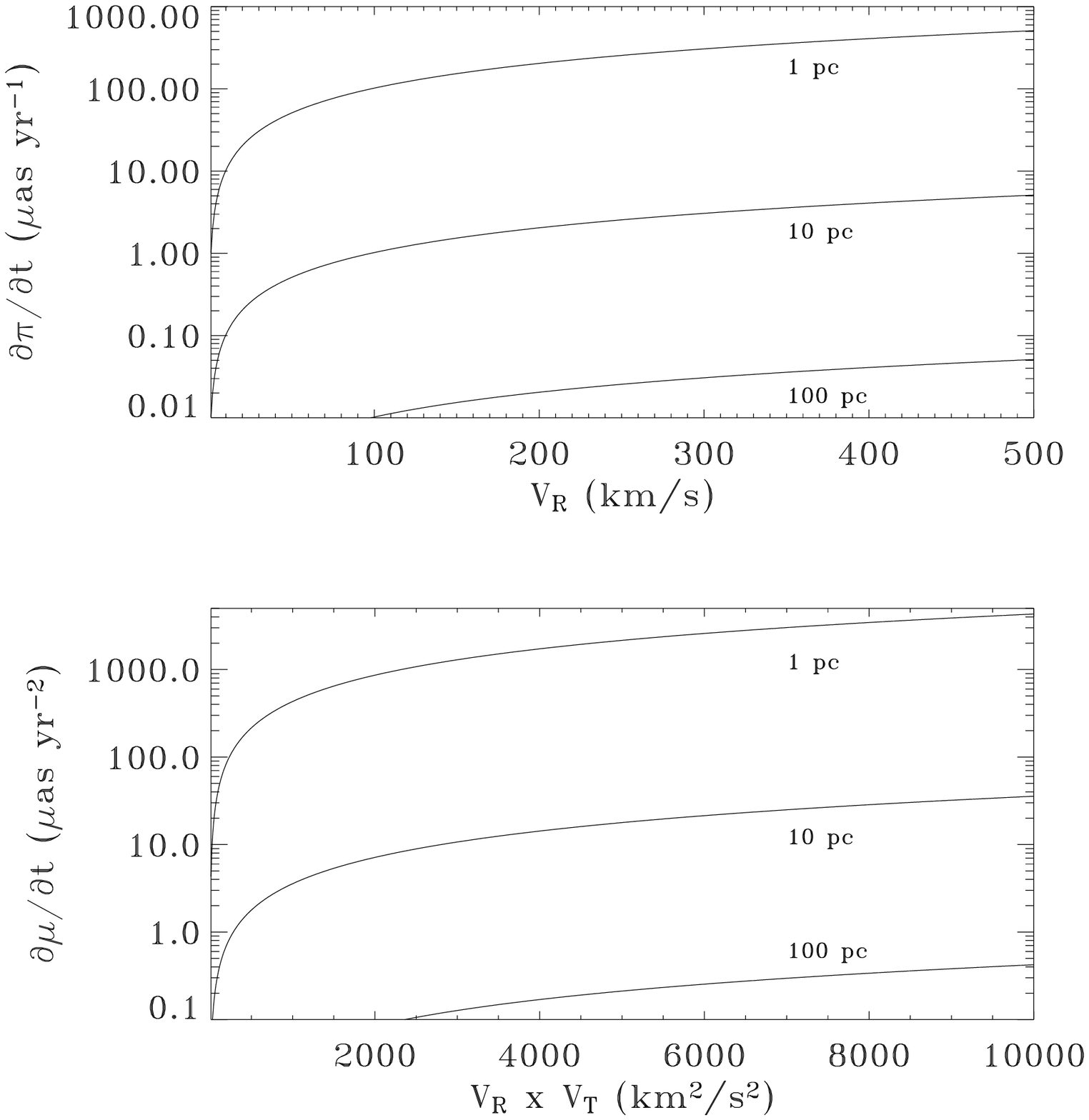}
\end{figure}

\clearpage

\begin{figure}
\epsscale{1.}
\plotone{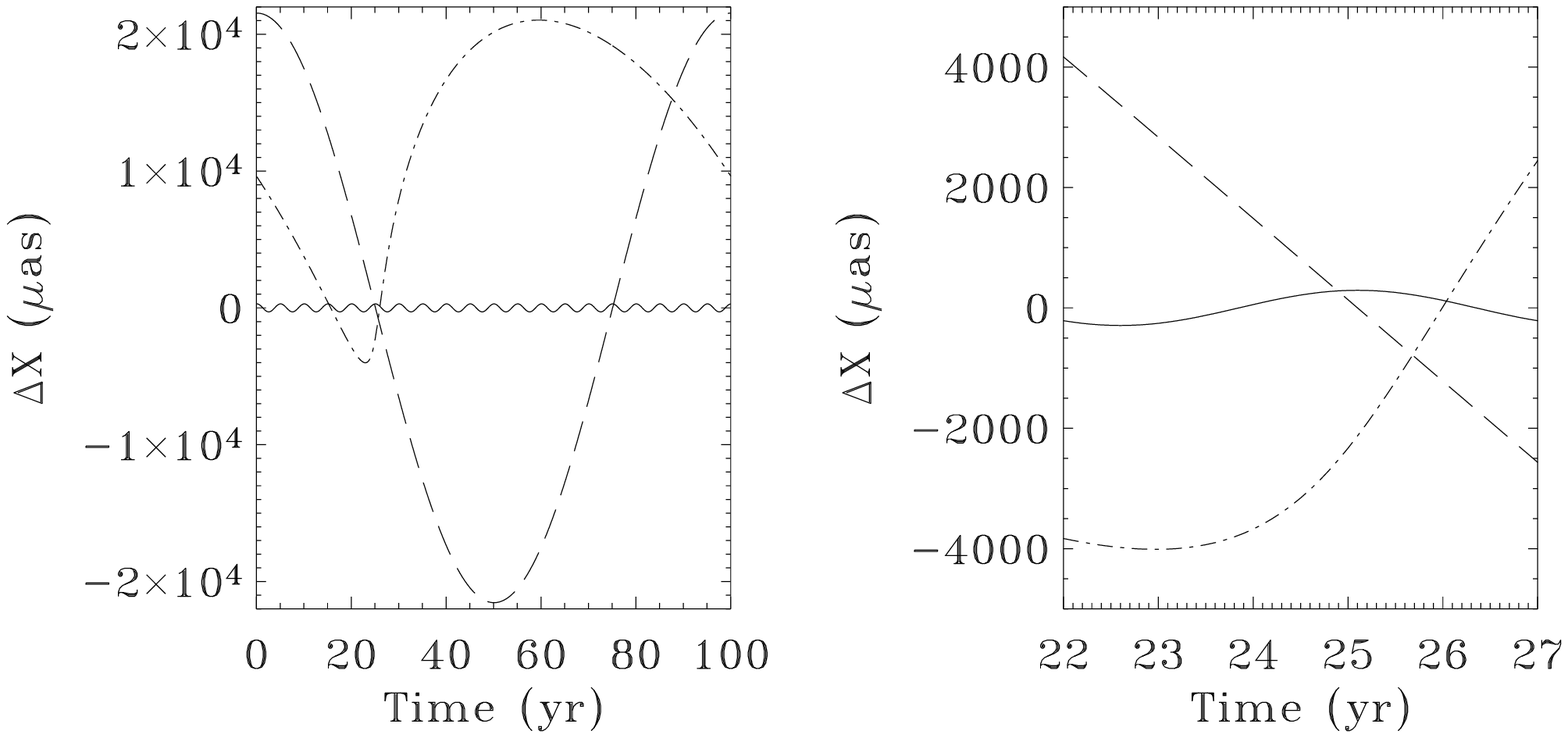}
\end{figure}

\clearpage

\begin{figure}
\epsscale{0.80}
\plotone{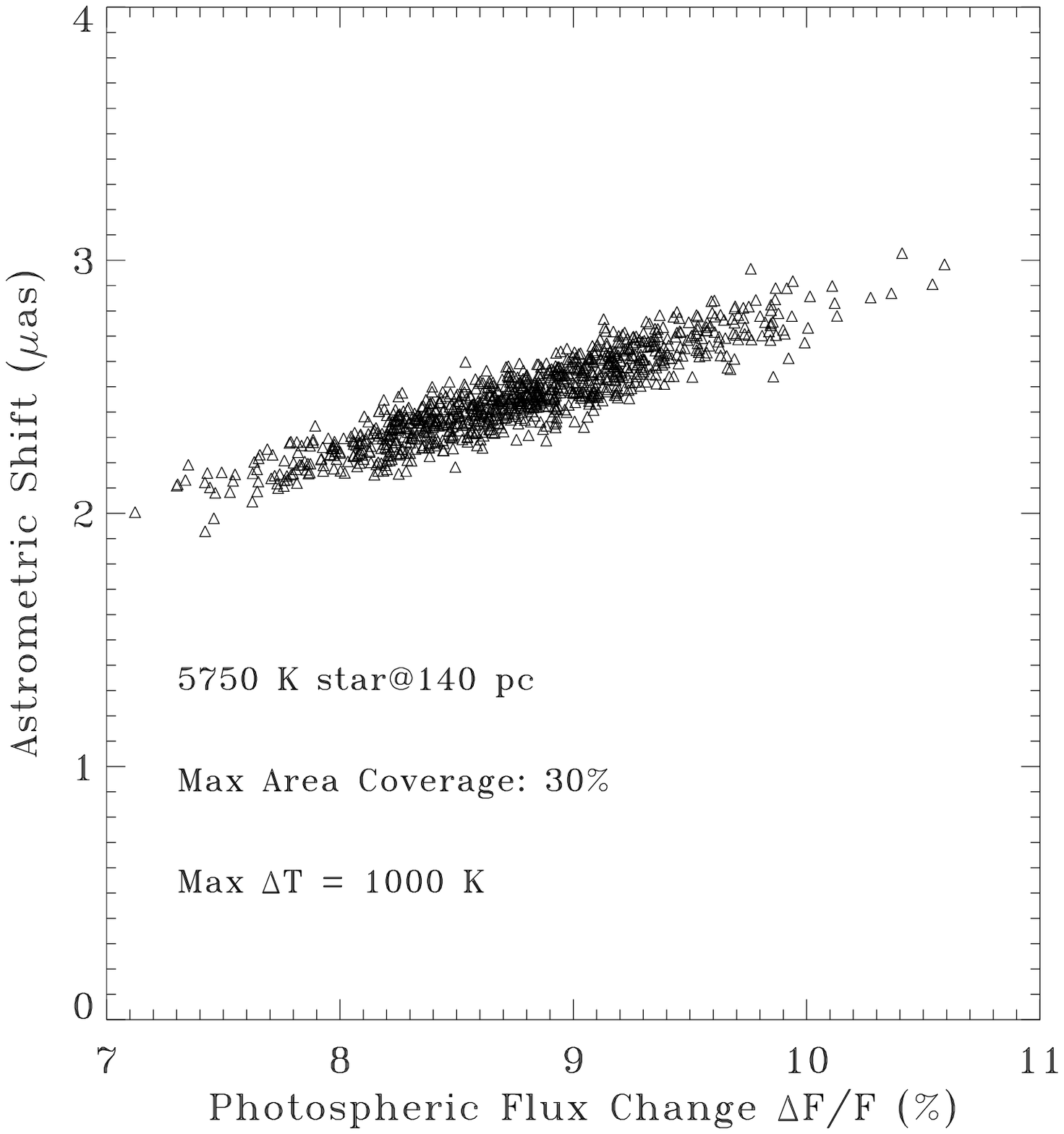}
\end{figure}

\clearpage

\begin{figure}
\epsscale{0.80}
\plotone{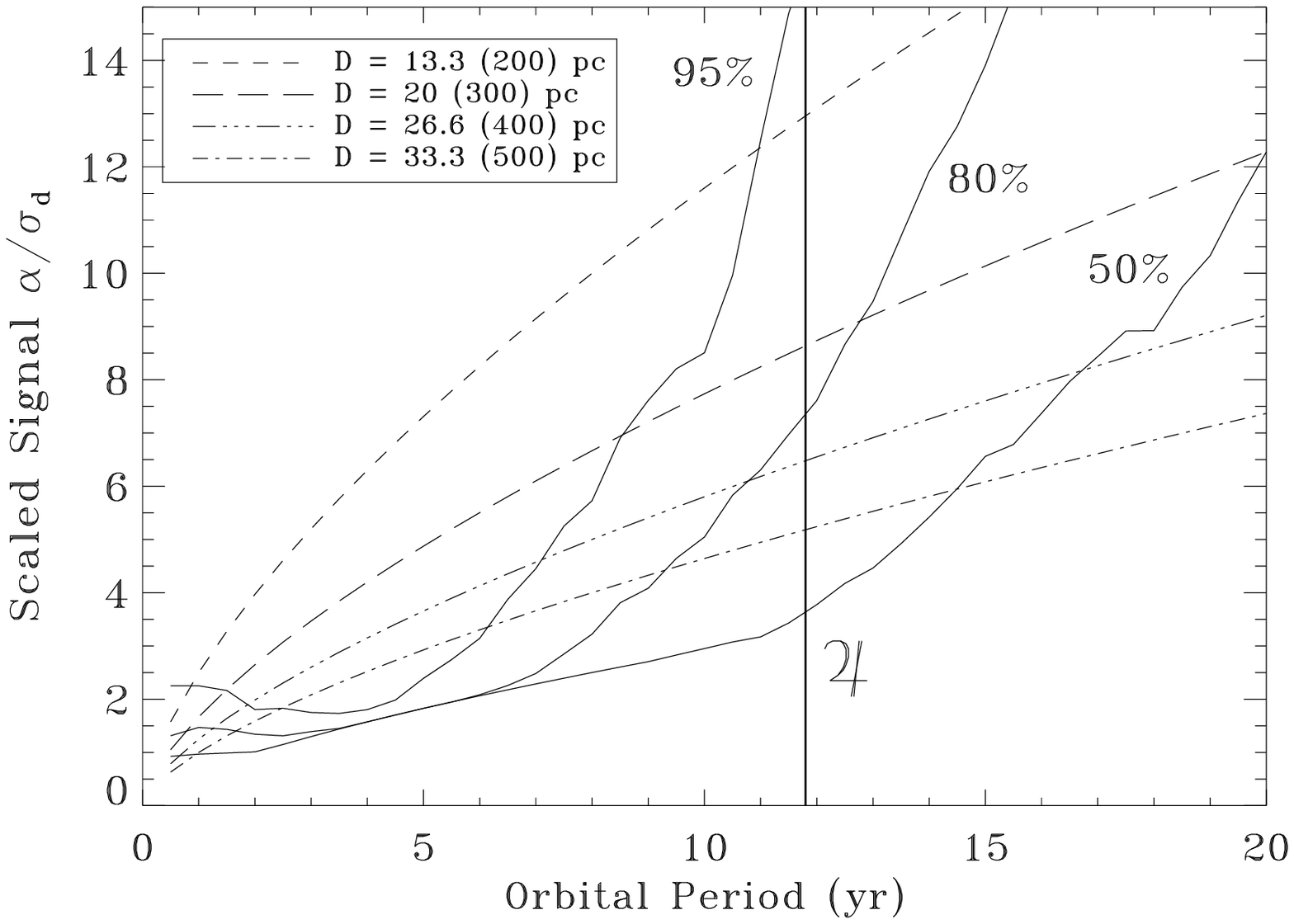}
\end{figure}

\clearpage

\begin{figure}
\epsscale{0.80}
\plotone{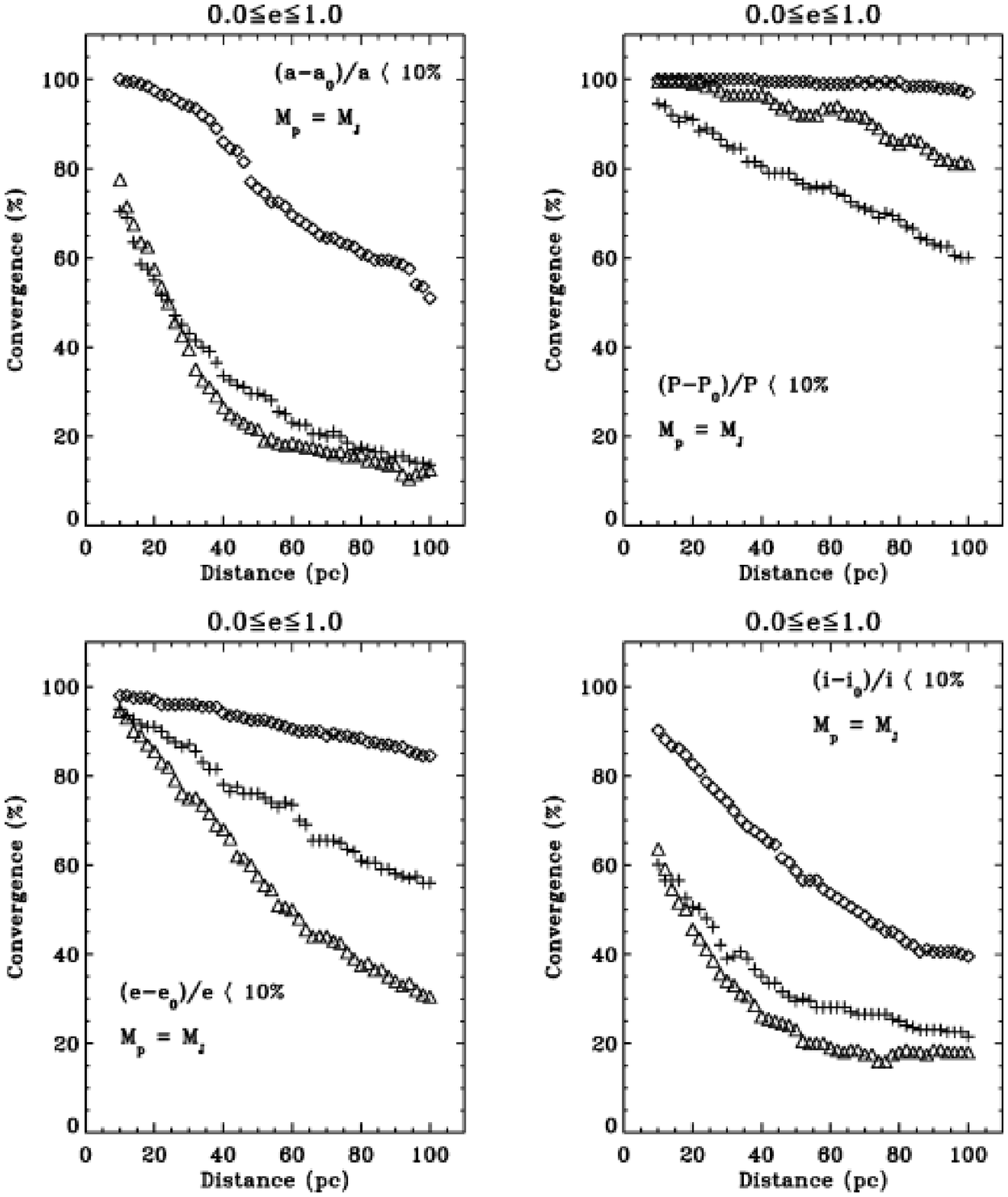}
\end{figure}

\clearpage

\begin{figure}
\epsscale{0.80}
\plotone{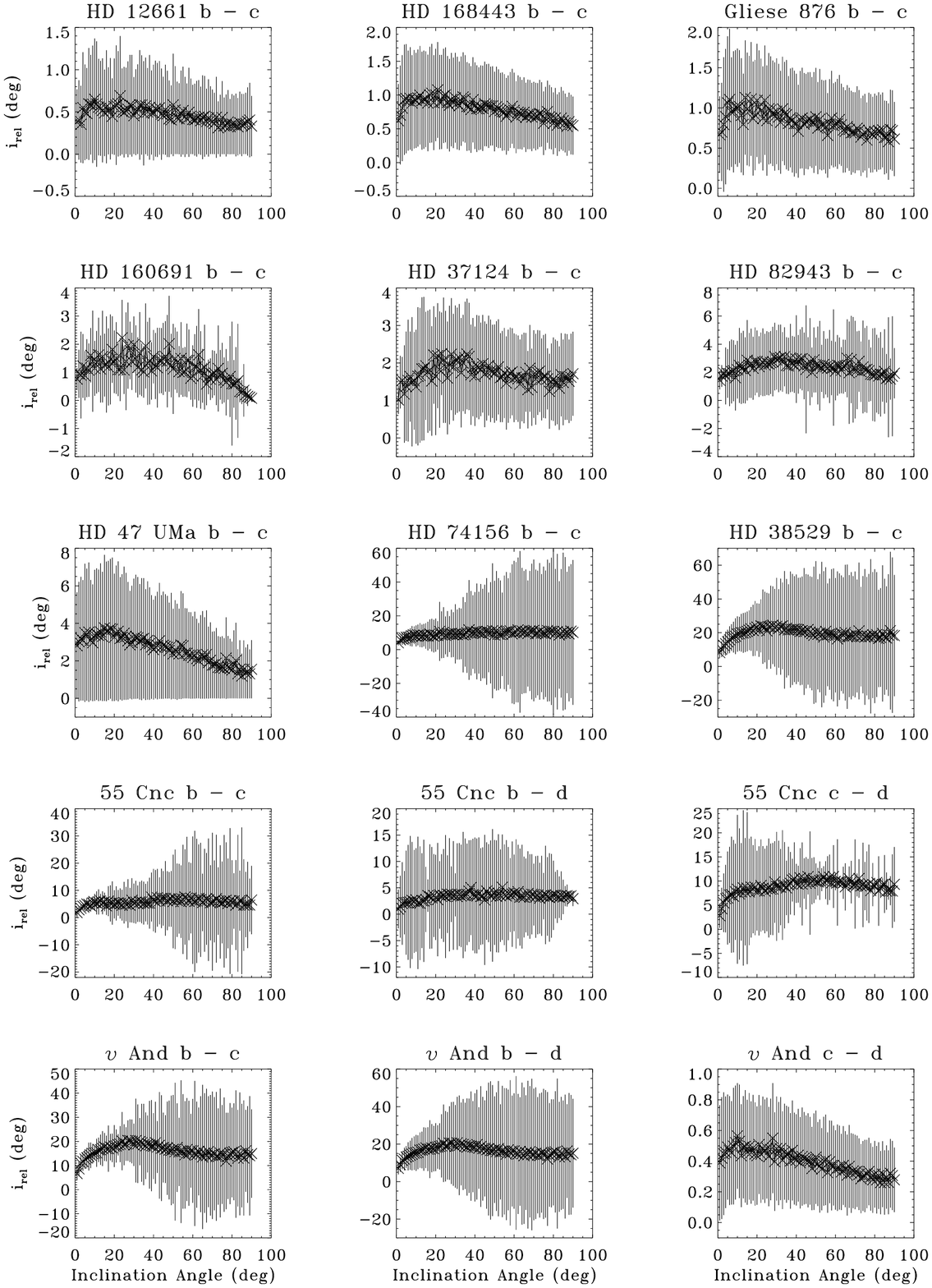}
\end{figure}

\clearpage

\begin{figure}
\epsscale{0.80}
\plotone{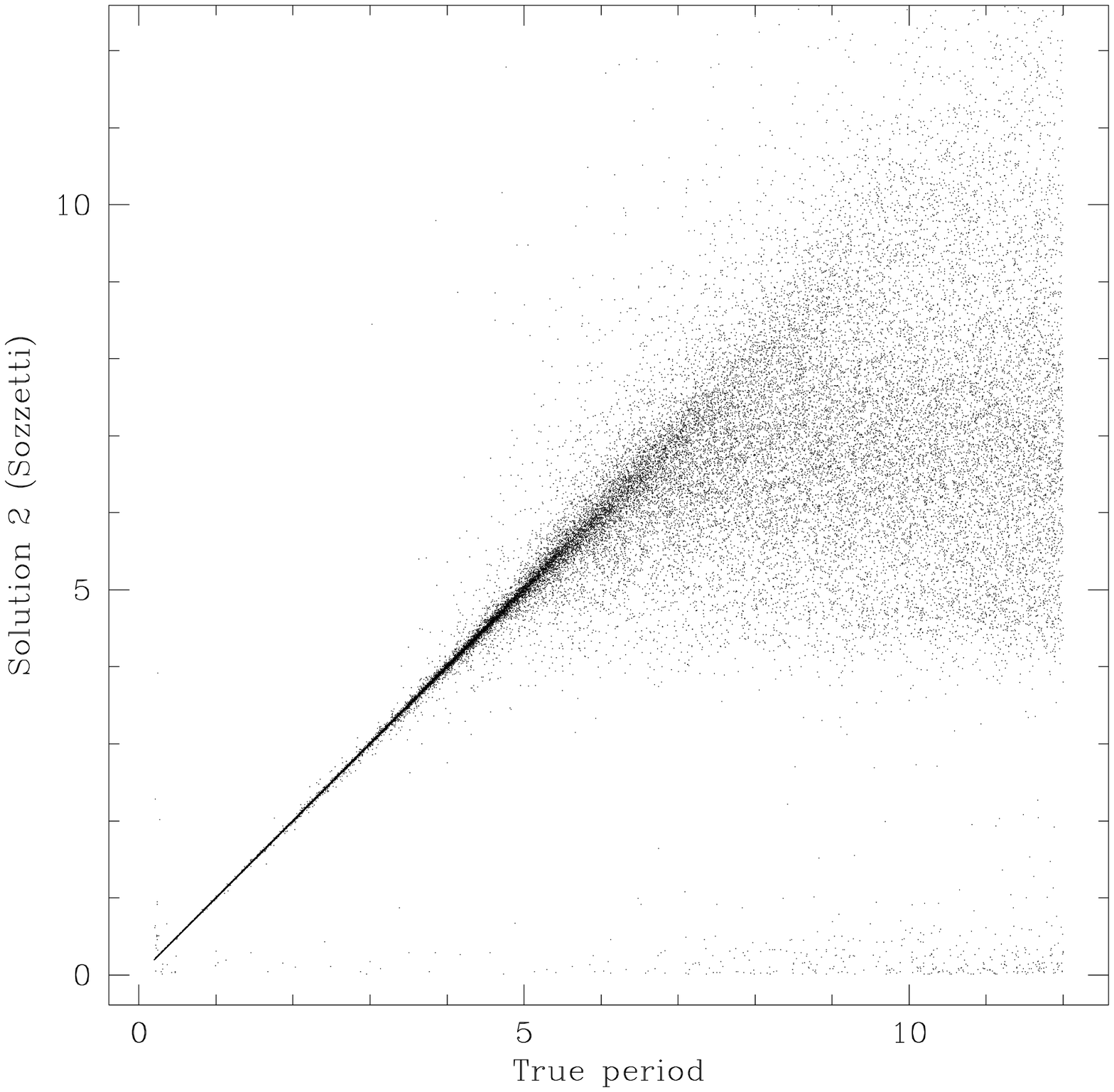}
\end{figure}

\clearpage

\begin{figure}
\epsscale{0.80}
\plotone{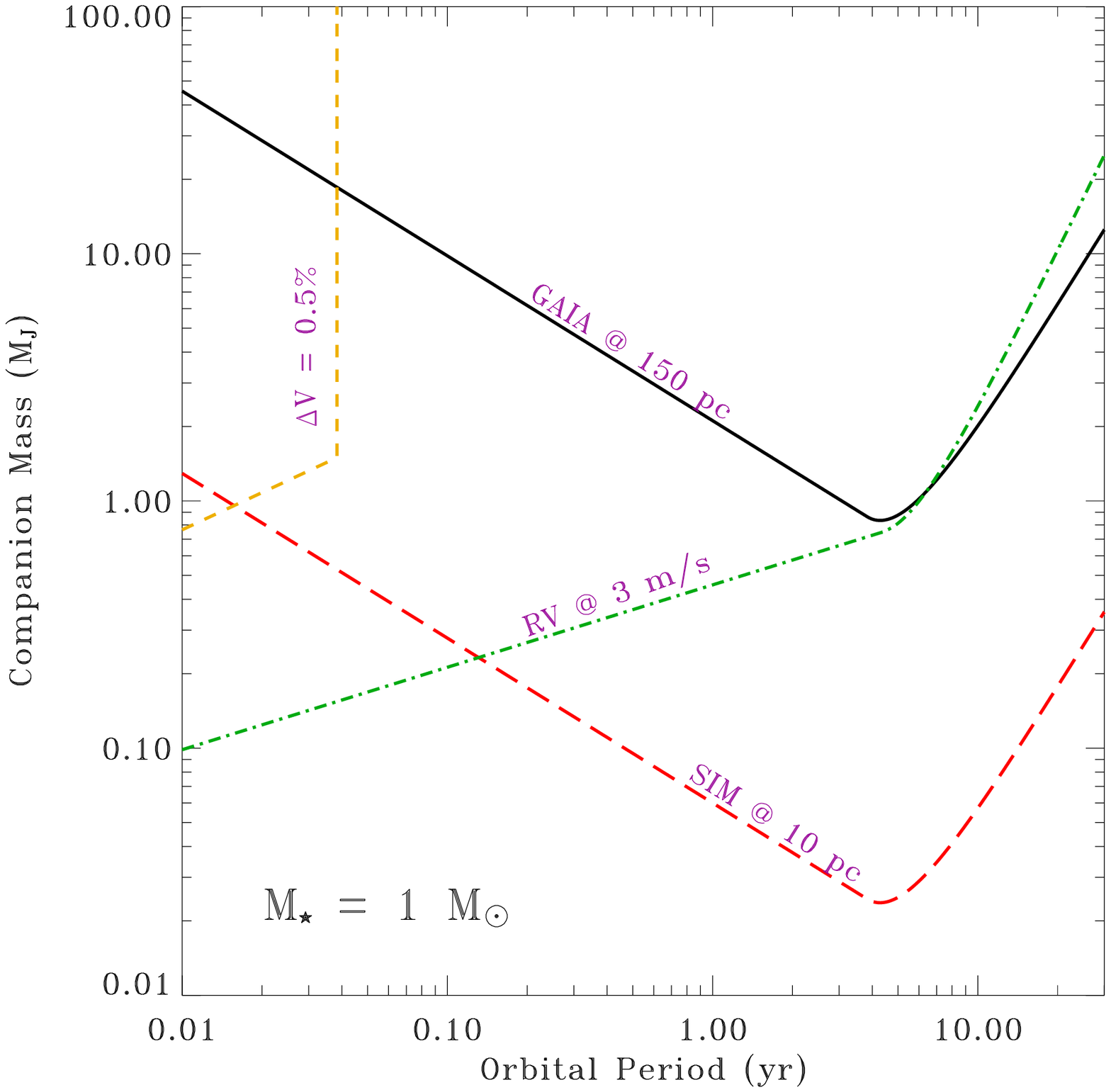}
\end{figure}

\end{document}